%% file: main.tex
\documentclass[nonacm,sigplan]{acmart}

\usepackage{color}
\usepackage{xcolor}
\usepackage{subfigure}
\usepackage{multirow}
\usepackage{xspace}
\usepackage{booktabs} 
\usepackage{listings} 
\usepackage{enumitem}
\setlist[enumerate]{leftmargin=*, parsep=0pt, itemsep=0pt, topsep=2pt}
\setlist[itemize]{leftmargin=*, parsep=0pt, itemsep=0pt, topsep=2pt}
\usepackage{comment}
\usepackage{graphicx}
\usepackage{textcomp}
\usepackage{caption}
\usepackage{algorithm}
\usepackage{algpseudocodex}
\usepackage{comment}

\newif\ifturnoffcomments
\ifturnoffcomments
    \newcommand{\jz}[1]{}
    \newcommand{\nej}[1]{}
    \newcommand{\reminder}[1]{}
\else
    \newcommand{\jz}[1]{{ \textcolor{blue} {#1}}}
    \newcommand{\nej}[1]{{ \bf \textcolor{red} {Natalie: #1} }}

    \newcommand{\reminder}[1]{ {\mbox{$<==$}} [[[ { \bf \textcolor{orange} {#1} } ]]] {\mbox{$==>$}}}
\fi
\newcommand{\ignore}[1]{}

\newcommand{\design}{\textsc{Arcus}\xspace}
\newcommand{\myparagraph}[1]{\vspace{0.04in}\noindent{\bf {#1}.}}
\newcommand{\mycircle}[1]{\raisebox{.5pt}{\textcircled{\raisebox{-.9pt} {#1}}}}

\begin{document}

\title{\design{}: SLO Management for Accelerators in the Cloud with Traffic Shaping}

\author{Jiechen Zhao$^{\varheartsuit}$, Ran Shu$^{\clubsuit}$, Katie Lim$^{\spadesuit}$, Zewen Fan$^{\diamondsuit}$, \\Thomas Anderson$^{\spadesuit}$, Mingyu Gao$^{\diamondsuit}$, Natalie Enright Jerger$^{\varheartsuit}$}
\affiliation{%
  \institution{$^{\varheartsuit}$ University of Toronto, $^{\clubsuit}$ Microsoft Research, $^{\spadesuit}$ University of Washington, $^{\diamondsuit}$ Tsinghua University}%
  \country{}%
}

\begin{abstract}

Cloud servers use accelerators for common tasks (e.g., encryption, compression, hashing) to improve CPU/GPU efficiency and overall performance. However, users' Service-level Objectives (SLOs) can be violated due to accelerator-related contention. The root cause is that existing solutions for accelerators only focus on isolation or fair allocation of compute and memory resources; they overlook the contention for communication-related resources. Specifically, three communication-induced challenges drive us to re-think the problem: (1) Accelerator traffic patterns are diverse, hard to predict, and mixed across users, (2) communication-related components lack effective low-level isolation mechanism to configure, and (3) computational heterogeneity of accelerators lead to unique relationships between the traffic mixture and the corresponding accelerator performance. The focus of this work is meeting SLOs in accelerator-rich systems. We present \design{}, treating accelerator SLO management as traffic management with proactive traffic shaping. We develop an SLO-aware protocol coupled with an offloaded interface on an architecture that supports precise and scalable traffic shaping. We guarantee accelerator SLO for various circumstances, with up to 45\% tail latency reduction and less than 1\% throughput variance. 

\end{abstract}

\maketitle 
\pagestyle{plain} 

\section{Introduction}\label{sec:intro}

To serve the growing demand for online software services~\cite{tail-at-scale}, CPU efficiency is increasingly crucial due to stagnating performance~\cite{esmaeilzadeh2011darksilicon}. Common tasks, such as security and data processing, can consume up to 82\% of CPU cycles in the cloud~\cite{kanev2015profiling,sriraman2020accelerometer,karandikar2023cdpu,karandikar2021grpcacceleration}, limiting resources for user applications. To mitigate this, clouds leverage \textit{hardware acceleration} to offload auxiliary tasks~\cite{karandikar2023cdpu,wolnikowski2021zerializer,kwon2020fvm,azuresmartnic,aws-aqua,taylor2020asiccloud}. Accelerators mainly bring two benefits: (1) saving CPU cycles, (2) improving the overall throughput that the system can sustain. 

A diversity of accelerators already exists in PCIe-attached I/O devices for providers to use, such as SmartNICs~\cite{amd-xilinx-smartnic,aws-nitro,liu2019ipipe,lin2020panic,pismenny2021autonomous}, DPUs/IPUs~\cite{galles2021pensando,intel-ipu,dpu-3}, GPUs~\cite{zhang2018gnet,silberstein2016gpunet,aws-gpu-video-encoding,gpu-nvcomp-compression}, or accelerator cards such as FPGAs~\cite{intel-ipu,shu2019dua,aws-aqua,aws-f1,chung2018brainwave,azuresmartnic} and ASICs~\cite{intel-quickassist,ranganathan2021google-video-accelerator,karandikar2023cdpu,azure-corsica-compression-asic}. Ideally, when incorporating accelerators with virtual machines (VMs), an accelerator SLO should be guaranteed so that user's end-to-end service target is not violated by the integrated accelerator. An SLO specifies two aspects: (1) a precise performance number and (2) and low variance~\cite{sutherland2020nebula,tail-at-scale}. Examples of accelerator SLO policies for a user can be (1) 100K IOPS of encryption as the accelerator SLO, in an encrypted database that offers a user 100K QPS under 99th percentile (99th\%) guarantee, (2) 5 Gbps of compression as the accelerator SLO, for RocksDB users whose SLO is set to be 5 Gbps under 99.9th\% guarantee. The focus of this paper is to guarantee accelerator SLO when integrating accelerators into the cloud. 

However, we find severe accelerator SLO violations with existing solutions, which primarily focus on fairness/isolation of compute and memory resources for accelerators~\cite{liu2019ipipe,khalilov2023osmosis,grant2020fairnic,lin2020panic}. They overlook contention on the communication-related resources (e.g., root complex, PCIe interconnects, buffers, and queues). As an example of showcasing the importance of communications, suppose $Y$ and $Z$ are the minimal ingress and egress rates required for an accelerator to satisfy the SLO. If (1) data are injected slower than $Y$, or (2) available bandwidth of the output channel is less than $Z$, accelerator SLOs will very likely be violated. Such communication-related problems hurt SLOs regardless of whether compute and memory allocation is proper or not within the accelerator sub-system. In reality, those cases happen frequently, e.g., (1) one user injecting too much traffic, overloading communication resources, (2) PCIe contention when running out of PCIe credits or root complex buffer space~\cite{tian2021cloudfpga-pcie-contention,neugebauer2018understandingpcie,vuppalapati2024understanding}, or (3) mis-arbitrating some users over others~\cite{yu2019ava}.

We find that violations of accelerator SLOs result from three fundamental causes related to communication. 

\begin{itemize}
    \item Traffic patterns are too diverse to predict. Even though an individual tenant is unlikely to be problematic, unpredictable contention depends on co-located traffic. Prior work only considers simple pattern mixtures~\cite{lin2020panic,liu2019ipipe,grant2020fairnic,kwon2021flexcsv,eran2022flexdriver,eran2019nica,ruan2019insider}, assuming well-shaped and same-sized. 
    \item Multiple communication-related components are possible contention sources, becoming sources of SLO violations. 
    \item Each accelerator has its own computational characteristics, leading to different relationships between traffic and accelerator performance. 
\end{itemize}


However, it is challenging to eliminate the above causes. First, each VM's traffic pattern is unknown a priori in the public cloud, let alone the traffic mixture from multiple VMs. Thus, avoiding SLO violations based on predictions or online scheduling is hard. Second, components have limited low-level isolation mechanisms, e.g., VMs’ traffic is not isolated across PCIe lanes but allocated by credits~\cite{neugebauer2018understandingpcie}. Therefore, we cannot rely on existing mechanisms to isolate traffic without re-designing the motherboard. Third, prior work treats accelerator I/Os the same as network I/Os or storage I/Os~\cite{grant2020fairnic,lin2020panic,eran2019nica,min2021gimbal,liu2019ipipe}, without heterogeneity awareness. 

To this end, our goal is to offer predictable accelerator performance and guarantee accelerator SLO. Our approach does not re-design CPU hardware or break the privacy boundary of VM traffic. We propose a communication-centric approach, with a protocol, an offloaded interface, and an architecture with the following two insights. 

\textbf{Insight 1}: The communication-induced SLO violation problem needs a communication-oriented solution. We propose to manage accelerator traffic by modeling them as flows and performing accurate traffic shaping to guarantee accelerator SLOs. Traffic shaping has been widely used in datacenter networks (e.g., in switches and NICs) for isolation enforcement~\cite{radhakrishnan2014senic,ballani2011oktopus,kumar2019picnic,angel2014pulsar}. It changes traffic from one pattern to another, e.g., limiting the rates and re-shaping message sizes. We consider the unpredictable accelerator flows transferring within a server through insufficiently isolated infrastructure as similar to the datacenter network setting. Therefore, the traffic shaping approach, in principle, is well-suited to addressing traffic-induced SLO violations in our context.

\begin{figure}
    \centering
    \includegraphics[width=0.99\columnwidth]{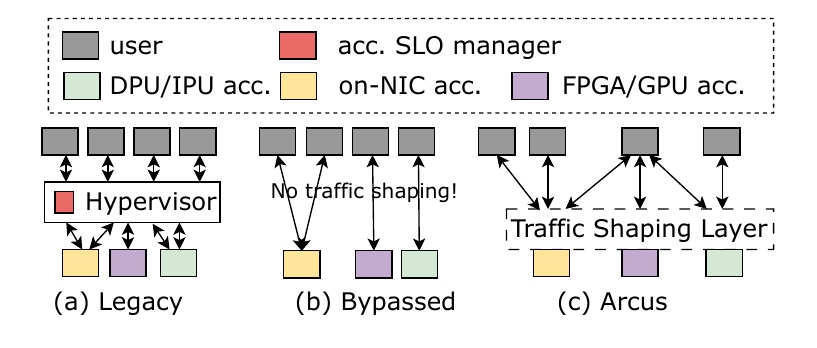}
    \caption{Accelerator management architectures. 
    }
    \label{fig:intro}
\end{figure}

\textbf{Insight 2}: Achieving good SLO management for accelerators requires an interposition layer to intercept all traffic. The legacy architecture can leverage the hypervisor as a good interposition layer (Fig.~\ref{fig:intro}(a)) which has visibility to re-shape any traffic~\cite{ji2023yama,zhang2022justitia,kumar2019picnic,klimovic2017reflex,agache2020firecracker}. Thus it mitigates the challenges mentioned above. However, such architecture still suffers from up to 25\% performance variance due to CPU interference (Sec.~\ref{eva:breakdown}). Worse, CPUs would become the performance bottleneck~\cite{kwon2020fvm,azuresmartnic}.\footnote{Handling I/O activities greater than 100+ Gbps consume tens of cores due to context switches, guest interrupts, etc.~\cite{kwon2020fvm,azuresmartnic}.} Alternatively, existing isolation/fairness-oriented approaches bypass the hypervisor (Fig.~\ref{fig:intro}(b)) for near-native accelerator performance~\cite{lin2020panic,grant2020fairnic,khalilov2023osmosis}. This architecture is not an SLO-oriented design due to (1) no heterogeneity awareness, (2) no communication contention avoidance, and (3) lack of interposition for management such as traffic shaping. Instead, we propose to proactively shape accelerator traffic on the accelerator side hardware, which gets rid of CPU interference by offloading. In addition, decisions are made separately by intercepting all traffic to a certain accelerator, with the awareness of heterogeneity and communication resource contention. Third, hardware-based traffic shaping enforces accurate SLO for each flow. Finally, our work enables the hardware to be programmable, thus this architecture can generalize accelerator SLO management for all circumstances, (1) under arbitrary traffic patterns, (2) under different usage modes, and (3) for arbitrary accelerators.

Based on the two insights, we present \design{}. A new SLO-aware protocol is realized on the architecture in Fig.~\ref{fig:intro}(c). We first learn the characteristics of heterogeneous compute and poorly-isolated communication infrastructure. Then we learn which traffic pattern mixture is more predictable for SLO management and which mixture leads to SLO violations more easily. A software runtime makes shaping decisions online for all accelerator flows. Finally, our protocol relies on a fast and accurate traffic shaping mechanism in hardware to re-shape incoming traffic on-the-fly on demand. 

We build our prototype in a host-FPGA system. \design{} introduces low overheads while reducing 99.9th\% latency by up to 45\% and keep throughput variances within 1\%. To evaluate end-to-end applications, we also build three prototypes where accelerators are located on FPGA-based SmartNICs, near-storage FPGAs, and PCIe-attached FPGAs. 
Results show 48--98\% of throughput variance reductions, and accelerators can deliver up to 1.43$\times$ overall throughput improvements and 58.9\% CPU cycle savings for RocksDB applications. 

\begin{itemize}
    \item We are the first to identify the importance of communications to address SLO violations of heterogeneous accelerators in the cloud. 
    \item We incorporate a hardware-based mechanism with runtime software as a novel SLO-aware protocol supporting the traffic shaping approach. The proposed architecture supports this protocol without changing CPU hardware or VM exposure abstraction. 
    \item The software runtime in the control plane runs algorithms based on learned characteristics with considerations of contention in insufficiently isolated communication infrastructure, and awareness of computational heterogeneity.  
    \item \design{}-enabled systems are SLO-oriented, in which SLO management just requires configuring a few SLO-related parameters rather than managing detailed resource allocations and low-level states. \design{} architecture and protocol generalize cases in realistic scenarios for various accelerator types. 
    \item \design{} virtually offloads part of the hypervisor functionality, i.e., traffic shaping-based SLO management, onto the device hardware. This work follows successful production-scale offloading examples in public clouds~\cite{azuresmartnic,aws-nitro}. 
    \item We integrate accelerators into various end-to-end applications with accelerator SLO guarantees. Thus, providers can improve CPU efficiency, overall throughput, and accelerator utilization. The predictability also enables flexible resource management such as over-subscription. 
\end{itemize}

\section{Background}\label{sec:background}

In this section, we emphasize the requirements of SLO-oriented designs (Sec~\ref{back:slo-policy}) and concepts in accelerator contexts from communication perspectives (Sec~\ref{back:paths}). 

\subsection{SLO-oriented Design Targets}\label{back:slo-policy}

Focusing on fairness and isolation is a different design goal from meeting SLO guarantees. Papers targeting better fairness and isolation for accelerators~\cite{khalilov2023osmosis,liu2019ipipe,grant2020fairnic} typically enforce allocation requirements on individual components, such as shared buffers or the accelerator themselves. However, designing systems for such targets is not sufficient for SLO management. Uniquely, preserving SLOs needs to (1) precisely guarantee a performance number and (2) under a certain availability~\cite{carvalho2014longtermSLO,tail-at-scale,sutherland2020nebula}. For instance, guaranteeing 40\% compute and 30\% memory resources to be well-isolated or fairly shared to a user does not directly lead to meeting a user's SLO in the contract saying ``\textit{the probability my VM gets at least $x$~Gb/s of computational power should be no smaller than 99\% over the next 3 months}.''~\cite{azure-managed-burst} Our paper dives into accelerator management from the SLO perspective. 

\subsection{Communication-centric SLO Management}\label{back:paths}

We focus on communication-induced SLO violations which have not been well-explored to date. In addition, we demonstrate the importance of communications, focusing on paths, traffic patterns, and accelerator heterogeneity. 

\myparagraph{Path} 
When invoking an accelerator, we define the logical flow of invocation streams as a path, architecturally representing data movement from one system component to another. Different paths may or may not share the same physical communication channel and physical interface. Paths build the foundation of this work's abstraction from communication perspectives. 

\begin{figure}
    \centering
    \includegraphics[width=0.99\columnwidth]{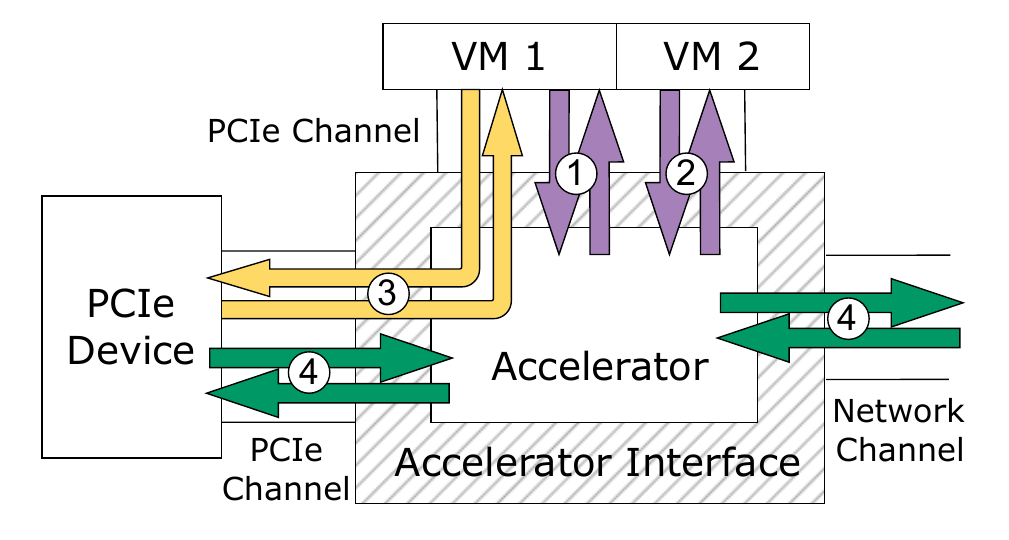}
    \caption{Possible paths to invoke an accelerator.
    }\label{fig:paths}
\end{figure}

\myparagraph{Path diversity}
There are many possible data paths for VM users to invoke an accelerator. Based on the accelerator interface, we classify the paths into three categories in Fig.~\ref{fig:paths}.
\begin{itemize}
    \item \textit{Function call mode}, where the VM's program triggers loopback traffic with a returned result \mycircle{1}, \mycircle{2}. 
    \item \textit{Inline mode (NIC)}, where the accelerator sits on the TX or RX path between the VM and the NIC device \mycircle{3}. 
    \item \textit{Inline mode (P2P)}, where the accelerator is on-path to/from a device (e.g., NICs, SSDs, GPUs) \mycircle{4}. 
\end{itemize}

\myparagraph{Diverse path combinations}
Different users may use the same path or different paths listed above. Each path can be contended by different VMs (i.e., \textit{inter-VM intra-path contention}), e.g., \mycircle{1} vs. \mycircle{2} and \mycircle{3} vs. \mycircle{4} in Fig.~\ref{fig:paths}. In each VM, multiple request streams from different services may contend on the same path (i.e., \textit{intra-VM intra-path contention}), e.g., \mycircle{1} vs. \mycircle{3}. Additionally, resource contention can happen across different paths if they share PCIe channels or device interfaces (i.e., \textit{inter-path contention}), e.g., \mycircle{2} vs. \mycircle{3}. 

\myparagraph{Diverse traffic pattern combinations}
An individual tenant’s traffic pattern can vary, including parameters such as message sizes and injection load. Further, multiple co-located users form a combination of mixed traffic patterns containing multiple heterogeneous streams.

\myparagraph{Heterogeneity of accelerators from communication viewpoints}
Each accelerator has its specialized logic and data flow. Our paper emphasizes their heterogeneity from a communication perspective. We call it ``non-linearity". Linearity represents read/write or send/receive, where egress and ingress bandwidth are equal, and larger messages linearly take higher bandwidth. This work considers the following non-linearity of accelerator traffic. 

\textit{Message sizes}. 
Each accelerator has a unique (non-linear) curve between compute throughput and its input data sizes. 

\textit{Egress/ingress bandwidth ratios}.
The bandwidth requirements of an accelerator can differ in terms of ingress and egress paths. 
There are several possible value ranges for the ratio $\frac{egressbw}{ingressbw}$ (denoted as $R$=$\frac{Eb}{Ib}$).  

\begin{itemize}
    \item $R$=1. For example, the output ciphertext of AES-256-CTR is always the same length as the input plaintext. 
    \item $R$>1, e.g., decompression falls into this category. 
    \item $R$<1, e.g., compression falls into this category.  
    \item $Eb$ is fixed. For example, SHA-3-512 has a fixed output message size of 64B, no matter how large the input is. 
\end{itemize}

\myparagraph{Basics of traffic shaping}
Rate limiting and changing the sizes of packets are the major techniques for traffic shaping in datacenter networks, enforcing isolation and providing precise shaping~\cite{radhakrishnan2014senic,ballani2011oktopus,kumar2019picnic,angel2014pulsar}. While those techniques are typically located in switches or NIC's TX path for outgoing traffic, our context is to limit the pace of fetching from DMA buffers for each accelerator-related path.

\begin{figure}[!b]
    \centering
    \includegraphics[width=0.99\columnwidth]{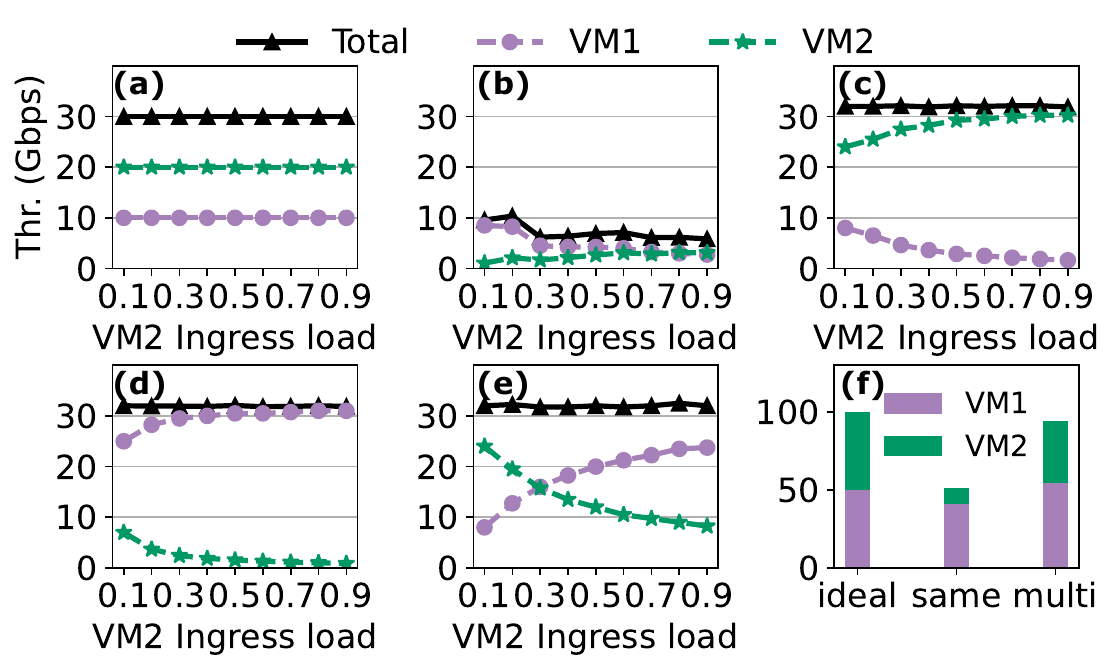}
    \caption{Representative results of case studies in Table~\ref{table:profiling-cases}.}
    \label{fig:panic-profiling}
\end{figure}

\section{Motivation \& Design Principles}\label{sec:motivation}

\subsection{Inaccurate Accelerator Resource Provisioning}\label{motivation:contention}

To motivate our design, we separately study the effects from (1) accelerator interface and (2) communication-related resources. 
We study if current systems can provision accelerator resources accurately based on SLOs. 
The metrics we parametrize are path diversity and traffic pattern mixtures. 
Our takeaways are summarized at the end of Sec.~\ref{motivation:contention}. 

\input{tables/profile-case-table}

\myparagraph{Accelerator-related inaccuracy} 
We select four cases to show the effect of traffic pattern combinations in a state-of-the-art system, PANIC~\cite{lin2020panic}. In all four cases, two VMs share an IPSec accelerator, whose overall capacity is 32Gbps at maximum for full load, MTU-sized packets; The IPSec's SLOs of VM1 and VM2 are set to be 10Gbps and 20Gbps, respectively. Accurate accelerator resource provisioning should guarantee the two VMs to be allocated throughput equal SLO. These four cases are meant to capture the accelerator interface's effect on SLO violation without the effect of communication-related resources. We use an on-FPGA traffic generator to tune injected traffic patterns from each VM. The patterns across CaseT$_{pattern1}$--CaseT$_{pattern4}$ are summarized in Table~\ref{table:profiling-cases}. 
Load means the injection rate by the traffic generator. 
The two VMs have the same priority.

Fig.~\ref{fig:panic-profiling}(a) shows the ideal SLO the two VMs should achieve. Compared to this, we make three observations based on the studies shown in Fig.~\ref{fig:panic-profiling}(b)-(e). First, the overall bandwidth of the IPSec accelerator is significantly affected by the mixture of message sizes. For instance, in Fig.~\ref{fig:panic-profiling}(b), CaseT$_{pattern1}$ only delivers 18--32\% throughput of the maximum (32Gbps). Second, not only are SLOs violated in all four cases, but the results also demonstrate a lack of fairness, i.e., a 50\%/50\% split. CaseT$_{pattern2}$ and CaseT$_{pattern3}$ never achieve a fair throughput allocation. CaseT$_{pattern1}$ only achieves fairness when input loads for VM1 and VM2 are 0.1 and 0.7, respectively. Similarly, for CaseT$_{pattern4}$, the fairness point is 0.1 and 0.3. Third, when ingress load rates increase, in CaseT$_{pattern4}$, VM1 gets varied throughput, 26--75\% of the maximum. In addition, when one VM's accelerator ingress load increases, the accelerator throughput of its co-located VM may decrease (e.g., CaseT$_{pattern1}$) or even increase (e.g., CaseT$_{pattern3}$).

\myparagraph{Communication-related inaccuracy} 
Next, we look at communication-related SLO violations. We put a traffic generator in each VM, where accelerator traffic contends for PCIe resources. To make sure the SLO violations we observe are all from communication-related contention, we duplicate the accelerator interface for each VM. Each VM owns a synthetic 50Gbps accelerator, a separate accelerator queue, and a separate DMA engine. We set the two VMs with the same priority, so ideally each VM should use their own accelerator at 50Gbps. We showcase the effect of path diversity. In CaseP$_{same\_path}$, two VMs share the same path, i.e., inline (NIC) RX path, while CaseP$_{multi\_path}$ involves different paths. The path and traffic patterns are summarized in Table~\ref{table:profiling-cases}. 

Fig.~\ref{fig:panic-profiling}(f) shows that, in CaseP$_{same\_path}$, even though there is no accelerator interface contention, VM1 gets 4$\times$ higher throughput than VM2 due to PCIe contention. In addition, in CaseP$_{same\_path}$ the overall throughput is only 55\% of that of CaseP$_{multi\_path}$; CaseP$_{multi\_path}$ achieves 85\% of the ideal throughput of PCIe Gen 3.0 x8. This is because CaseP$_{multi\_path}$ takes advantage of the full-duplex feature of PCIe (host-to-FPGA and FPGA-to-host directions), while CaseP$_{same\_path}$ has two VMs both contending only in the FPGA-to-host direction. Though omitted in Fig.~\ref{fig:panic-profiling}(f), we find that CaseP$_{same\_path}$ has an 85\% drop on the overall throughput. The difference between CaseP$_{same\_path}$ and CaseP$_{multi\_path}$ implies that simultaneously managing accelerator traffic from diverse paths improves the accelerator's overall throughput, compared with all VMs contending a single path.

\myparagraph{Takeaways}
(1) For each accelerator type, the injection rates and message sizes of each stream should be considered or even under control. Otherwise, the uncertainty of traffic patterns leads to significant inaccuracy of accelerator provisioning. Traffic shaping is well-suited to addressing such limitations. 
(2) Due to the lack of low-level isolation mechanisms on communication-related resources, traffic shaping needs the understanding of how they are contended under different traffic patterns and path combinations. Directly augmenting such resources with a new hardware isolation mechanism would not work due to significant hardware changes. 

Note that the studies showcased in Sec.\ref{motivation:contention} are \textit{not aimed to motivate certain heuristics} for a certain case. Selected studies are representative results to demonstrate how challenging and complex SLO management will be if solutions strive to fit each combination case-by-case. These results \textbf{aim to motivate us to rethink} a unified approach, i.e., traffic shaping, that can fundamentally handle any case.

\subsection{High Complexity of SLO Management}\label{motivation:complexity}

\myparagraph{Architectural limitations on effective traffic shaping}
Although traffic shaping is effective in hypervisor-managed systems~\cite{ji2023yama,zhang2022justitia,kumar2019picnic}, the hypervisor itself can become the performance bottleneck~\cite{kwon2020fvm,azuresmartnic} and lead to SLO violations. Meanwhile, hypervisor-bypassed systems have no mediation layer to intercept, monitor, and shape traffic between accelerators and VMs. Thus, previous in-hypervisor traffic shaping approaches cannot be used in state-of-art systems. 

One naive way is to offload traffic shaping from the hypervisor onto the accelerator side. Common traffic shaping usually requires schemes coordinating with the rest of the system ~\cite{kumar2019picnic,klimovic2017reflex,agache2020firecracker}. However, there is no protocol yet to manage how offloaded traffic shaping interacts with the host software and the VMs.

\myparagraph{No SLO managers with global visibility across all paths} 
Current managers for NIC accelerators~\cite{liu2019ipipe,grant2020fairnic,lin2020panic,khalilov2023osmosis} or discrete cards~\cite{kwon2021flexcsv,intel-quickassist} are designed ad hoc and only enable visibility for one path. Constructing per-path SLO managers in a distributed manner will lose global visibility and suffer from coordination overheads. For instance, deciding which traffic pattern to shape needs global synchronization across different managers because they own SLO and traffic states separately. For another example, flows either from the same or different paths may contend for PCIe resources. 

To offer a centralized SLO manager for all paths, the key challenge is how to build it with low complexity and good scalability. The centralization comes with more states to keep and manage across all paths and traffic. Conventional wisdom of accelerator managers cannot scale well (1) with accelerators' throughput increasing, (2) with user numbers increasing. This results from their tracking low-level states and making low-level resource allocation decisions~\cite{khalilov2023osmosis,liu2019ipipe}.

\subsection{Design Principles}

To achieve our goals, we advocate rethinking the system architecture to guarantee accelerator SLOs. Our proposal follows a few principles we set out to address challenges described in Sec.~\ref{motivation:contention} and Sec.~\ref{motivation:complexity}.

\begin{itemize}
    \item \textbf{Accelerator flow abstraction}. We manage accelerator-related traffic with a flow abstraction, similar to flows for network traffic. Each VM can trigger multiple accelerator flows, and each physical channel can sustain multiple flows. Flows can be uni- or bidirectional. This abstraction enables generalized traffic management for all paths, usage modes, and contention cases. 
    \item \textbf{A centralized \& offloaded interface.} With the flow abstraction, we offer an interface to intercept, monitor, and manage all accelerator traffic from all paths. We offload it onto accelerator side hardware to avoid host resource contention and achieve high performance. This interface has a global interposition for all paths and flows for each accelerator. Thus, we eliminate the need to coordinate across different managers.
    \item \textbf{Per-flow SLO-aware traffic shaping mechanism}. Unlike previous SLO managers who reactively schedule accelerator time slices to each VM, we proactively shape the traffic pattern of each flow. This is an SLO-oriented mechanism designed in \design{} interface, which directly takes SLO-related parameters. 
    \item \textbf{Decoupled protocol supporting  traffic shaping}. We design a decoupled protocol to coordinate the mechanism with the VMs. \design{} interface mediates and re-shapes \design{}-to-accelerator traffic. VM-to-\design{} protocol is unchanged for compatibility. The decoupling enables \design{} to proactively shape accelerator traffic. 
    \item \textbf{Profiling-assisted traffic shaping}. We propose to perform offline profiling to learn $Capacity(t,X,N)$, i.e., the available capacity of an accelerator $X$ at a given time $t$ shared by $N$ VMs, w.r.t. traffic patterns $T$, path mode combinations $P$, and system settings $S$ (e.g., PCIe bandwidth). We store this as a table for the control plane to make online decisions about traffic shaping. 
\end{itemize}

\section{\design{} Design}\label{sec:design}

\begin{figure}
    \centering
    \includegraphics[width=0.99\columnwidth]{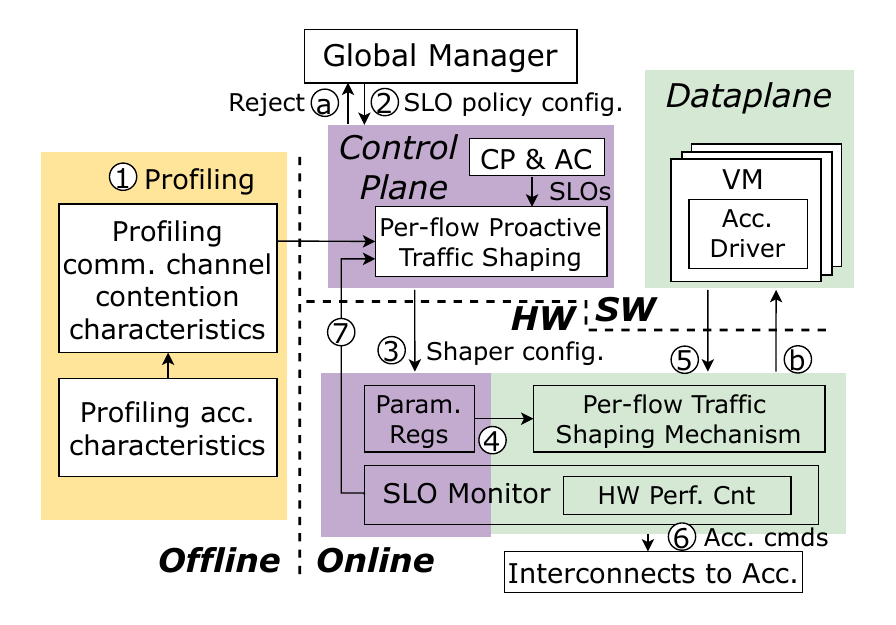}
    \caption{Workflow of an \design{}-enabled system. CP: capacity planning, AC: admission control. }
    \label{fig:workflow}
\end{figure}

Fig.~\ref{fig:workflow} shows a high-level overview of the entire \design{}-enabled system. The purple and green regions represent data (Sec.~\ref{design:dataplane}) and control planes (Sec.~\ref{design:slo-management}). The software runtime handles control plane operations. First, \design{} maintains a profile table, which stores the results of $Capacity(t,X,N)$ learned from offline profiling \mycircle{1}. The software runtime also takes in SLOs from different users \mycircle{2}. It then selects a path to the accelerator required, considering factors such as contention, and decides what traffic pattern to shape for a new user. The decisions are passed to the hardware parameter registers \mycircle{3}, which configure the underlying hardware mechanism \mycircle{4} implementing the traffic shaping (Sec.~\ref{design:mechanism}). After decision-making and configuration finishes, VMs can trigger accelerator invocations on the dataplane \mycircle{5}, bypassing the host software for near-native performance. The invocations are translated into commands and sent to the interconnects for accelerator access \mycircle{6}. A back-pressure mechanism assists the traffic shaping \mycircle{b}. During the entire service, the control plane keeps track of hardware counters for each flow's performance \mycircle{7} to monitor SLO metrics. If a new shaping decision is made on a flow, steps \mycircle{3}-\mycircle{4} repeat without interrupting dataplane operations. 

Integrating \design{} into existing systems requires the following modifications. First, runtime software should be installed in each client server that hosts accelerator-enabled VMs. Second, a new \design{} interface should be added with (1) a register file and performance counters accessible to the runtime and (2) a programmable hardware mechanism for traffic shaping. The protocol between VMs and accelerators still remains as before, preserving SR-IOV-like near-native performance. VM users do not need to manage underlying resources (e.g., buffers, queues, accelerators) as in prior work~\cite{khalilov2023osmosis,liu2019ipipe}.

\subsection{Dataplane Protocol}\label{design:dataplane}

\input{design-sections/protocol}

\subsection{Per-flow Traffic Shaping Mechanism}\label{design:mechanism}

\input{design-sections/rate-limit}

\subsection{Control Plane: SLO Management Runtime}\label{design:slo-management}

\input{design-sections/slo-management}

\section{Evaluation}\label{sec:eva}

We use methodologies in Sec.~\ref{eva:method} to answer the following. 

\begin{itemize}
    \item Can \design{} maintain accelerator SLOs compared to baselines? Can \design{} realize near-zero variance for throughput and tail latency and accurate SLOs (Sec.~\ref{eva:slo})? 
    \item How well does \design{} perform compared to the ideal and how much overheads are added for each flow (Sec.~\ref{eva:per-flow})? 
    \item What characteristics are learned offline (per-flow accelerator heterogeneity in Sec.~\ref{eva:per-flow} and communication-related contention in Sec.~\ref{eva:multi-flow})? How well does the control plane make decisions dynamically based on such results?
    \item How scalable is \design{} across many flows (Sec.~\ref{eva:multi-flow})? 
    \item How effective is \design{} in an end-to-end application setting, e.g., secure key-value stores, encrypted live migration, near-storage computation, and checksum and compression in RocksDB (Sec.~\ref{eva:ete})? 
\end{itemize}

\subsection{Methodology}\label{eva:method}

\input{experiment-sections/testbed}

\subsection{Achieving SLO Guarantee}\label{eva:slo}

We separately evaluate the SLO about (1) performance accuracy and (2) variance.

\input{experiment-sections/ev-slo}

\subsection{Understanding \design-enabled System Deeply}\label{eva:breakdown}

We perform detailed analysis within \design{}. Sec.~\ref{eva:per-flow} introduces per-flow studies, including traffic shaping breakdown results, accelerator profiling for each flow, and dynamism when reconfiguration happens for each flow. Sec.~\ref{eva:multi-flow} studies different scenarios where multiple flows exist, including scalability, control plane decision-making processes, and concrete use cases of \design{}.

\input{experiment-sections/ev-breakdown}

\subsection{End-to-End Evaluation}\label{eva:ete}

\input{experiment-sections/ev-ete}

\section{Discussions}\label{design:rationale}

\input{other-sections/discussions}

\section{Related Work}\label{sec:related-work}

\input{other-sections/related-work}

\input{tables/prior-art}

\section{Conclusions}

This work presents \design{}, a protocol, and an architecture that guarantees accelerator SLOs in the cloud. 
We identify communication-induced SLO violations and address the problem based on a traffic shaping based system, incorporating a software runtime and a hardware mechanism together, with learned characteristics applied to traffic shaping decisions on the fly. 
Our FPGA-based prototypes eliminate throughput variances, reduce tail latency and improve system efficiency. 
Our protocol works for diverse accelerators used within different data paths, fitting across diverse traffic patterns.

\bibliographystyle{plain}
\bibliography{references}

\end{document}

%% file: tables/profile-case-table.tex
\begin{table*}
\centering\footnotesize
\caption{Settings of case studies shown in Fig.~\ref{fig:panic-profiling} to analyze accelerator provisioning (Sec.~\ref{motivation:contention}) in current systems. 
}

\begin{tabular}{ccccc}
\toprule
      & Accelerator Type                                                                                                   & Traffic Pattern Combination                                 & Path Combination                                & Ideal Capacity Planning                                                                                         \\
\midrule
CaseT$_{pattern1}$ & \multirow{4}{*}{\begin{tabular}[c]{@{}c@{}}Two VMs \\share one \\32Gbps IPSec\end{tabular}}                        & VM1 \{256B, load=0.1\}, VM2 \{64B, load=0.1-0.9\}             & N/A                                              & \multirow{4}{*}{\begin{tabular}[c]{@{}c@{}}Overall 30Gbps, \\SLO(VM1)=10Gbps, \\SLO(VM2)=20Gbps~\end{tabular}}  \\
CaseT$_{pattern2}$ &                                                                                                                    & VM1 \{256B, load=0.1\}, VM2 \{512B, load=0.1-0.9\}            & N/A                                              &                                                                                                                 \\
CaseT$_{pattern3}$ &                                                                                                                    & VM1 \{128B, load=0.1\}, VM2 \{512B, load=0.1-0.9\}            & N/A                                              &                                                                                                                 \\
CaseT$_{pattern4}$ &                                                                                                                    & VM1 \{1500B, load=0.1\}, VM2 \{512B, load=0.1-0.9\}           & N/A                                              &                                                                                                                 \\
\midrule
CaseP$_{same\_path}$ & \multirow{2}{*}{\begin{tabular}[c]{@{}c@{}}Allocate a separate \\50Gbps synthetic acc. \\to each VM\end{tabular}} & VM1 \{4KB, load=0.4\}, VM2 \{64B, load=0.1-0.9\}        & \begin{tabular}[c]{@{}c@{}}VM1 \{Inline (NIC) RX\}, \\VM2 \{Inline (NIC) RX\}\end{tabular} & \multirow{2}{*}{\begin{tabular}[c]{@{}c@{}}Overall 100Gbps, \\SLO(VM1)=50Gbps, \\SLO(VM2)=50Gbps\end{tabular}}  \\
CaseP$_{multi\_path}$ &                                                                                                                    & VM1 \{4KB, load=0.4\}, VM2 \{64B, load=0.1-0.9\}        & \begin{tabular}[c]{@{}c@{}}VM1 \{Function Call\}, \\VM2 \{Inline (NIC) RX\}\end{tabular}   &    \\     
\bottomrule
\end{tabular}
\label{table:profiling-cases}
\end{table*}

%% file: design-sections/protocol.tex
\begin{figure}
    \centering
    \includegraphics[width=0.99\columnwidth]{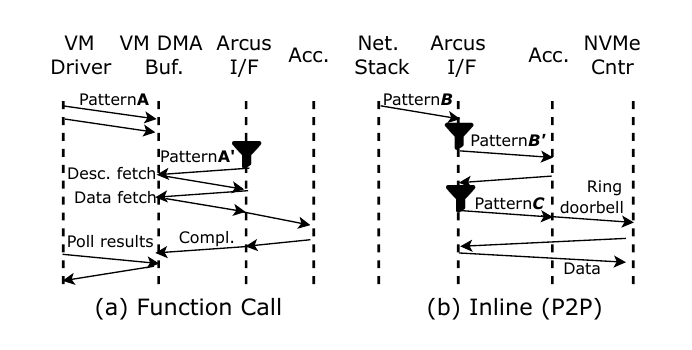}
    \caption{\design{} dataplane protocol for two modes of accelerator invocation. The funnel icon means traffic shaping. 
    }
    \label{fig:protocol}
\end{figure}

\myparagraph{Function call mode}
The protocol for this mode is illustrated in Figure~\ref{fig:protocol}(a). The accelerator driver installed in the VM pre-allocates a memory region, which can be DMA'd by the accelerator. The driver pushes its traffic to the local accelerator by placing payloads to be processed into the shared DMA buffer. Its traffic pattern is supposed to be \verb|PatternA|. Instead of accepting doorbell rings reactively, \design{} interface proactively fetches descriptors from the DMA buffer in \verb|PatternA'|, different from \verb|PatternA|. This new pattern is differentiated from VM-determined \verb|PatternA| and is decided by the runtime software. The decision-making process of \verb|PatternA'| is discussed in Sec.~\ref{design:slo-management}. The metadata inside the descriptor includes accelerator type, traffic pattern, and a pointer to the DMA buffer. After interpreting the descriptors created by the driver, \design{} fetches payloads and streams them to the local accelerator for computation. The completion writes back to the dedicated memory region. The driver polls the results of computation as the response of an accelerator invocation in function call mode. All fetching operations are DMA reads, and the completion is a DMA write.

In essence, our protocol \textit{decouples the dependency between invocations' submission rates} (determined by VMs) and their \textit{actual rates received by accelerators}. It moves the decision-making for accelerator traffic patterns from untrusted VMs to infrastructure-managed \design{} interface. This ``rate transformation" is performed on-the-fly, proactively creating SLO-friendly DMA traffic. An arbitrary number of hardware queues in \design{} interface can be allocated and bound to software queues, which are created by the driver for applications to interact with accelerators.

\myparagraph{Inline mode (P2P)} 
As Fig.~\ref{fig:protocol}(b) shows, our protocol also supports mediating storage I/Os for accelerator invocations. After receiving a packet from the network whose ingress traffic pattern is supposed to be under \verb|PatternB|, \design{} interface shapes the traffic towards NVMe controllers to be \verb|PatternB'|. Similarly, \verb|PatternB'| is determined by the runtime. The network packet headers are interpreted and translated into the descriptor that a local accelerator can digest and compute. The computed result is sent back to \design{} interface, which shapes another runtime-decided \verb|PatternC| to inject into NVMe sub-systems. Shaped traffic are packed as NVMe commands processed. The NVMe protocol over PCIe includes doorbell rings and data fetches, which are DMA reads to the buffer in \design{} interface. 

\myparagraph{Inline mode (NIC)} 
Most SmartNIC systems support this mode, making accelerators on-path~\cite{eran2019nica,eran2022flexdriver,liquidio,liu2019ipipe} or off-path~\cite{chen2024dpa,wei2023characterizing,liu2019ipipe}. On the TX path in this mode, most parts of the protocol are the same as in function call mode, except that the computed result would not be returned to the host alongside with the completion message. Instead, the result will be stored in packets and sent out. On the RX path, \design{} interface drains the on-NIC receive buffer in pull-based fashion. The fetching request pattern is shaped by \design{}.

%% file: design-sections/rate-limit.tex
We cannot rely on VM users to shape themselves because they are untrusted. Also, we find that realizing traffic shaping with high accuracy is challenging in the host software. For example, to stably limit a rate of 100Gbps, the software needs to pace 1KB accelerator I/O message every $\sim$80ns on average. Even high-resolution timers in today's software cannot guarantee such accuracy.

Our work chooses to pair a hardware-based rate limiter with each per-flow queue within \design{} interface. Our design provides near-ideal accuracy thanks to its cycle-level timers in hardware. Additionally, our design removes host resource contention related to traffic shaping. For a tractable rate limiter hardware design, there are two important aspects to consider: how friendly the algorithm is to hardware implementation and how easy it is to program at runtime.

\myparagraph{Algorithm}
We use a token bucket algorithm~\cite{token-bucket} because it is hardware-efficient, burstiness-friendly, and accurate. We tried a sliding window log algorithm, which is accurate by adding caches, but it is complex and memory-inefficient to implement. We considered fixed window counter and leaky bucket algorithms as they are resource efficient but they are not suitable for bursty request patterns. The token bucket algorithm needs two parameters, one more than the above alternatives. \verb|Bkt|$\_$\verb|Size| is the bucket size and \verb|Refill|$\_$\verb|Rate| is the refill rate to the bucket. Increasing \verb|Bkt|$\_$\verb|Size| means that a larger burst is permitted. The value of \verb|Refill|$\_$\verb|Rate| determines how often a token is added to the bucket. 

Our design supports two modes, shaping Gbps or IOPS. The only difference is to increase and decrease tokens based on the number of bytes, or the number of messages. We use hardware timers to perform token calculation every \verb|Interval| cycles. This delivers accurate calculations, not affected by CPU interference. Messages can be re-sized by splitting the payloads and duplicating another message header. 

\myparagraph{Programming interface}
The two parameters of the token bucket are stored in two separate registers, which are exposed to the host software as PCIe BAR address. The runtime software can use a memory-mapped I/O (MMIO) interface to directly read/write the two parameters at runtime.

%% file: design-sections/slo-management.tex
\input{algorithm/slo-mgmt-algo}

\myparagraph{Offline preparation} There are two static data structures prepared by the runtime. First, for each accelerator, it keeps an \verb|AccTable| entry, recording its optional path(s). For example, accelerator A is available located on server X. This entry records \verb|ServerXIPAddr|:\verb|PCIAddr|. Second, each entry of \verb|ProfileTable| records the pointer to profiled results of $Capacity(t,X,N)$. For each accelerator, all contention cases are swept and recorded in different entries. Each entry also has a 1-bit tag, representing if a traffic pattern combination on a path combination is \verb|SLO|-\verb|Friendly| or \verb|SLO|-\verb|Violating|. The former category is the candidate for the control plane to select for the next shaping configuration, while the latter category should be strictly avoided. 

\myparagraph{Capacity planning}
Allocating, re-allocating, and monitoring accelerator usage involves interactions between the runtime and the mechanism. The key data structure to support such interactions is a \verb|PerFlowStatusTable|. It is a dynamically updated table, indexed by \verb|FlowID|. Each entry includes the following information corresponding to each \verb|FlowID|: the VM ID, path ID and accelerator ID for this flow, per-flow SLO, the mechanism parameters configured for this flow, and the current SLO status measured from hardware counters.  

With \verb|PerFlowStatusTable|, our system supports capacity planning in the following three scenarios.

\begin{itemize}
    \item \verb|Scenario1|: Availability check. The runtime first reads \verb|ProfileTable| and checks the total capacity an accelerator can achieve in a particular context (pattern combination and path combination). Then the last column of \verb|PerFlowStatusTable| is read when the runtime calculates how much available capacity is left for such a context. 
    \item \verb|Scenario2|: New registration. A new \verb|PerFlowStatusTable| entry is created if the runtime makes sure the SLO target can be satisfied after checking the availability under a certain context. 
    \item \verb|Scenario3|: Runtime adjustment. In a certain entry, path ID and mechanism parameters can be reconfigured, if SLO targets are predicted to be violated. 
\end{itemize}

\myparagraph{Runtime algorithm}
As shown in Algorithm~\ref{alg:algorithm}, the runtime runs in each client server. It periodically checks if SLO is violated for each \verb|FlowID| (\verb|line4|). The capacity planning triggers \verb|Scenario3| if SLO violations start to occur. With a PathSelection() function, if the current path is overloaded while a new path is available, the runtime increases accelerator capacity for this flow to avoid SLO violations (\verb|line18|). ReshapeDecision() function finds appropriate mechanism parameters and the runtime configures them to the hardware (\verb|line20|-\verb|21|). After reconfiguration is committed, the runtime updates corresponding columns including path ID (if adjusted), mechanism parameters, and new SLO status (\verb|line6|). When a new flow comes, the runtime first checks the availability of the requested accelerator (i.e., \verb|Scenario1|). The runtime rejects this new registration if not enough availability left (\verb|line9|), corresponding to \mycircle{a} Reject in Fig.~\ref{fig:workflow}. Otherwise, \verb|Scenario2| is triggered, and \verb|PerFlowStatusTable| registers a new entry.

%% file: algorithm/slo-mgmt-algo.tex
\begin{algorithm}[t]
\caption{\design{} Runtime Algorithm.}
\label{alg:algorithm}
\begin{algorithmic}[1]
  \Procedure{Arcus Accelerator SLO Manager}{}
    \LComment{Run by every client server periodically}
    \For{each FlowID}
    \If{\textsc{SLOViolationChecker}()==FALSE} 
        \State \textsc{ReAdjustPattern}()
        \State Update \verb|PerFlowStatusTable|
    \EndIf
    \EndFor
    \While{OnNewRegist==TRUE}
        \If{ \textsc{!AdmissionControl}(policy, target)} 
            \State Reject registration
        \EndIf
        \State \textsc{CapacityPlanning}(NEW,policy,target)
    \EndWhile
  \EndProcedure

  \Function{SLOViolationChecker}{}
      \If{ReadSLOPerfCnts[FlowID] < target[FlowID]} 
        \State Return FALSE
      \EndIf
  \EndFunction

  \Function{AdmissionControl}{}
    \If{\textsc{CapacityPlanning}(CHECK)==FALSE} 
        \State Return FALSE
    \EndIf
  \EndFunction

  \Function{ReAdjustPattern}{}
    \If{ PathSelection() returns \textit{newPath}} 
        \State \textsc{CapacityPlanning}(ADJUST,\textit{newPath})
    \EndIf
    \State \verb|Param| $\gets$ ReshapeDecision()
    \State Config. \verb|Param| into \verb|TrafficShapingParamRegs|
  \EndFunction
\end{algorithmic}
\end{algorithm}

%% file: experiment-sections/testbed.tex
The results in Sec.~\ref{eva:slo}--\ref{eva:breakdown} come from a host-FPGA prototype where \design{} is implemented on the FPGA. The host CPU has dual 16-core Intel Xeon E5 2698v3 CPU sockets and 256GB RAM, running Ubuntu 18.04 with Linux kernel version 4.17.12 installed. The FPGA is connected to the CPU through a PCIe root complex with a PCIe Gen 3.0 x8 interface. Each FPGA has two 50Gbps Ethernet ports. 

We implement \design{} in System Verilog. The FPGA runs at 250 MHz, the same clock frequency as PCIe HIP. The datapath is 256 bits wide. Our prototype uses the built-in DMA engine on the FPGA. A PCIe interface encodes and decodes PCIe TLP headers when sending and receiving DMA messages. We augment the DMA engine with an external wrapper with SR-IOV arbiter (a simple round robin policy~\cite{intel-quickassist}) and queues similar to prior work~\cite{zazo2015pciedma-40gbpsfpga}, which in our case contains accelerator per-flow contexts (including queues and a rate limiter). The host sets control registers to tune parameters on the FPGA through an MMIO interface. VMs run on the CPU socket that attaches to the FPGA.

\myparagraph{Configurations} The following systems are our baselines. 
\begin{itemize}
    \item Host$_{no\_TS}$. The host connects with the FPGA, running a kernel-bypass library to access on-FPGA accelerators. The FPGA uses a weighted round-robin to arbitrate traffic across different users. No traffic shaping is supported. 
    \item Host$_{TS\_firecraker}$ and Host$_{TS\_reflex}$ are based on on-host software traffic shaping approaches in ReFlex~\cite{klimovic2017reflex} and FireCracker~\cite{agache2020firecracker}. Neither communication-related contention nor heterogeneity are considered, even though the average ingress rate can be rate limited on the host. 
    \item Bypassed$_{no\_TS\_panic}$. The host connects with the FPGA, bypassing the hypervisor. On-FPGA accelerators are accessed with a user-level library.  PANIC~\cite{lin2020panic} is the accelerator interface, with priority-based and weighted fair queuing policies. No traffic shaping is supported. There is no proactive SLO management and no considerations of heterogeneity and communication-related contention. 
\end{itemize}

\myparagraph{Paths}
Sec.~\ref{eva:slo} mainly studies function call mode, and inter-VM intra-path contention. Sec.~\ref{eva:breakdown} extends the explorations to intra-VM intra-path and inter-VM inter-path contention. Sec.~\ref{eva:ete} evaluates representative end-to-end applications that cover all categories.

%% file: experiment-sections/ev-slo.tex
\input{./tables/rate-limiter-config-table}

\myparagraph{High accuracy of guaranteeing performance numbers}
Table~\ref{table:rate-limiter-config-table} shows how the mechanism shapes an accelerator flow from 1Gbps to 1,000Gbps. This is fine-grained enough for single-digit Gbps services and sufficient to catch up with line rates of upcoming PCIe 5.0~\cite{adatapcie5,samsungpcie5} and 1Tbps Ethernet links~\cite{ethernet-roadmap}. Table~\ref{table:rate-limiter-config-table} shows the required values of two parameters (\verb|Bkt|$\_$\verb|Size| and \verb|Refill|$\_$\verb|Rate|). \verb|Interval| is the number of cycles to start another refill. To find the right pair quickly, we first fix \verb|Bkt|$\_$\verb|Size| to be a certain value, and then sweep \verb|Refill|$\_$\verb|Rate|. Even for 1,000Gbps SLO, \verb|Interval| is only 64 FPGA cycles (256ns), long enough to be easily implemented. This benefits from large \verb|Bkt|$\_$\verb|Size|, allowing the outcome insensitive to large bursts and message size variations.

\myparagraph{Low variance of performance}
We compare \design{}'s performance stability with Host$_{TS\_firecraker}$ and Host$_{TS\_reflex}$.

\textbf{\textit{Throughput}}.
In both baselines and \design{}, two users simultaneously send 4KB random read requests to the SSD. As for SLOs, we set SLO$_{user1}$ and SLO$_{user2}$ to be 300K IOPS and 200K IOPS. Both SLOs should be under 99th\% guarantee. We sample the throughput of the two users every 500 requests.

\begin{figure}[t]
  \centering 
  \subfigure[User1 (SLO=300K KIOPS)]{
      \includegraphics[width=0.7\columnwidth]{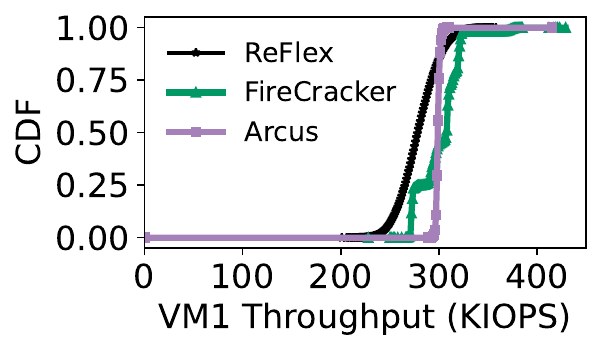}}
  \subfigure[User2 (SLO=200K KIOPS)]{
      \includegraphics[width=0.7\columnwidth]{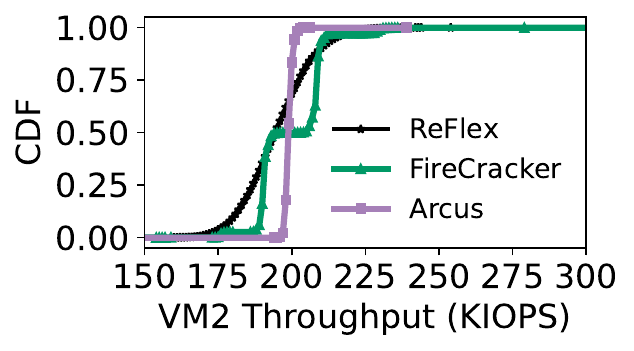}
  }
  \caption{CDF of throughput achieved by two users. CDF curve should be as sharp as possible for low variance. 
  }
  \label{fig:low-variance}
\end{figure}

Fig.~\ref{fig:low-variance} presents the CDF of throughput. The goal is to have the CDF curve as sharp as possible to represent low variance. \design{} achieves SLO$_{user1}$ and SLO$_{user2}$ numbers accurately. However, the two baselines suffer from CPU interference. The CPU processing of VMs leads to imprecise software token buckets and software timers and unpredictable execution times in software.

\textbf{\textit{Tail latency}}.
For ReFlex, 95th\%, 99th\%, and 99.9th\% latency are 128$\mu$s, 193$\mu$s and 299$\mu$s. For \design{}, 95th\%, 99th\%, and 99.9th\% latency are 104$\mu$s, 133$\mu$s and 162$\mu$s. The latency reductions for \design{} are 18.75\%, 31.09\%, and 45.82\%. The latency reduction comes from near-zero CPU usage and resultant CPU resource interference between software runtime, VM1, and VM2. 

\begin{figure*}[htbp]
  \centering
  \begin{minipage}{0.99\linewidth}
    \centering
    \subfigure[Profiling accelerators]{
    \includegraphics[width=0.32\columnwidth]{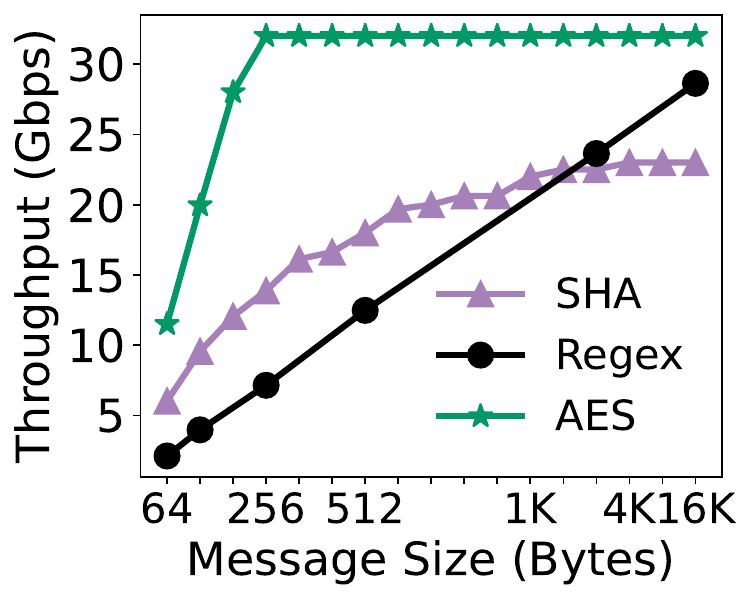}}
    \subfigure[Scaling flow numbers]{
    \includegraphics[width=0.32\columnwidth]{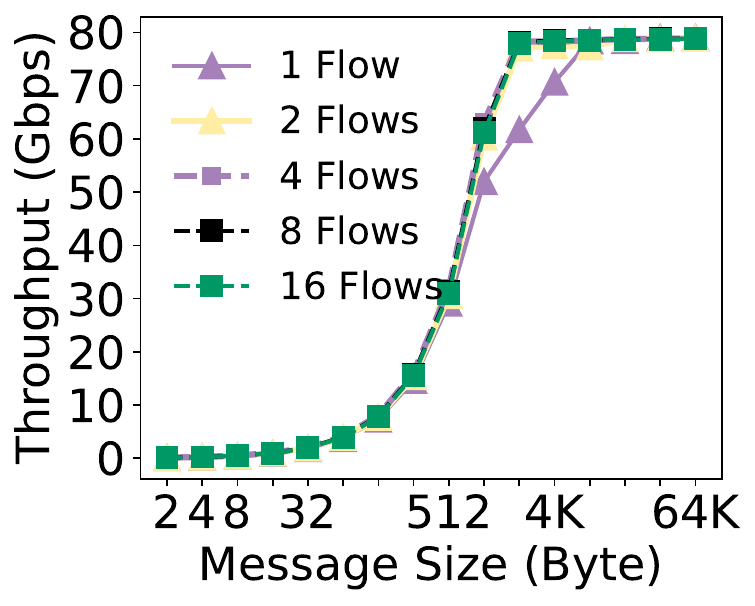}
    }
    \subfigure[16 RX flows vs. 4 TX flows ]{
    \includegraphics[width=0.32\columnwidth]{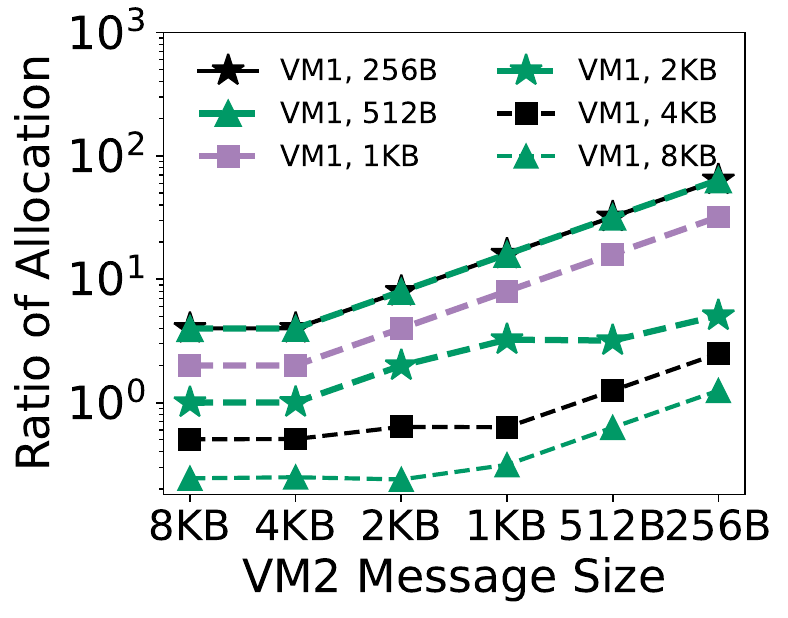}
    }
    \caption{Contention studies with combined factors (traffic pattern and direction, flow numbers, accelerator heterogeneity, etc.). }
    \label{fig:profiling}
  \end{minipage}\quad
\end{figure*}

\begin{figure}
  \centering
  \includegraphics[width=0.9\columnwidth]{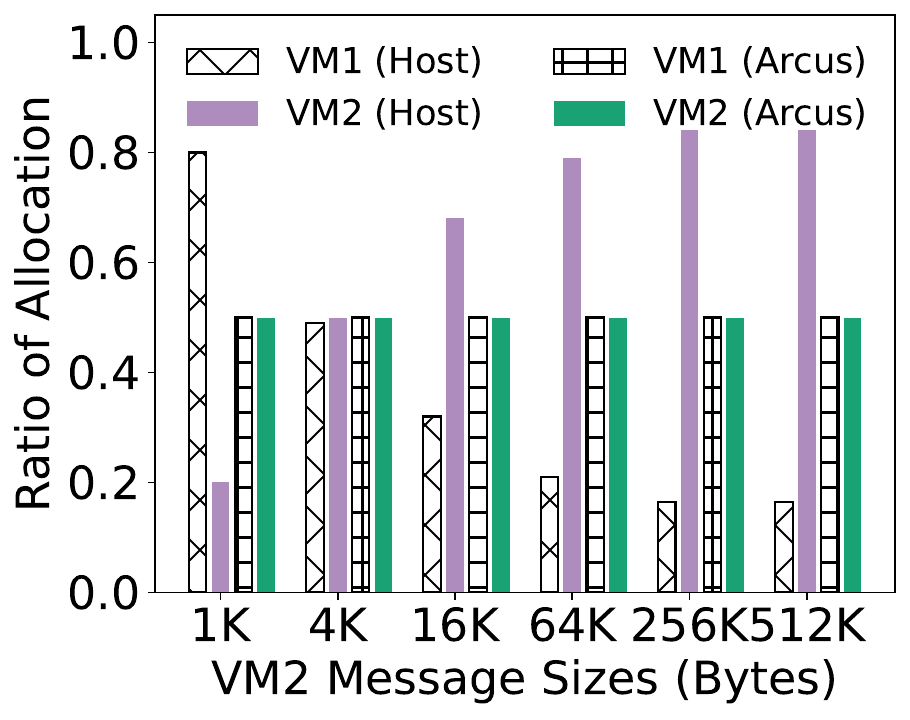}
  \caption{Use case1, SLO guarantee for large message users. }
     \label{fig:large-msg}
\end{figure}

%% file: tables/rate-limiter-config-table.tex
\begin{table}
    \centering\footnotesize
    \caption{Parameter table for accurate traffic shaping. }
    \begin{tabular}{cccc}
    \toprule
    \textbf{\textbf{SLO}} & 
    \begin{tabular}[c]{@{}c@{}}\textbf{Refill$\_$Rate} \\(\# tokens per interval)\end{tabular} & \begin{tabular}[c]{@{}c@{}}\textbf{Bkt$\_$Size} \\(\# tokens)\end{tabular} & \textbf{\textbf{Interval}}  \\
    \midrule
    1 Gbps                & 1,024                       & 512                   & 1000 cycles                 \\
    10 Gbps               & 4,096                      & 4,096                 & 800 cycles                  \\
    100 Gbps              & 16,384                     & 65,536                & 320 cycles                  \\
    1000 Gbps             & 32,768                     & 1,048,576             & 64 cycles                   \\
    \bottomrule
    \end{tabular}
\label{table:rate-limiter-config-table}
\end{table}

%% file: experiment-sections/ev-breakdown.tex
\subsubsection{Per-flow Studies}\label{eva:per-flow} 

We break down per-flow traffic shaping details and learn per-flow accelerator characteristics. 

We first study \myparagraph{Traffic shaping breakdown}
In the experiment of Fig.~\ref{fig:low-variance}, the latency cost of the traffic shaping mechanism is only 36ns. In comparison, software-based traffic shaping in Host$_{TS\_firecraker}$ and Host$_{TS\_reflex}$ takes $>$10$\mu$s.

Table~\ref{table:rate-limit-comparison} shows that our 25th, 50th, 75th, and 99th percentile throughput are all $\leq$1\%  from the SLO for VM1 in the experiment of Fig.~\ref{fig:low-variance}. We omit similar results for VM2 due to space limits. On the other hand, Host$_{TS\_firecraker}$ and Host$_{TS\_reflex}$ suffer from 6.5--11.7\% throughput loss and 8.7--24.3\% throughput over-provisioning at 25th and 99th percentile throughput. Such variations are harmful to SLO-oriented designs.

We also test synthetic accelerators with different cycles and compute time distributions such as Bi-modal, Poison, and Uniform. We guarantee SLOs for them at an optimal parameter setting. The results are omitted due to space.

\myparagraph{Accelerator heterogeneity} 
We choose three representative curve types, logarithmically, exponentially, and uniquely ad-hoc, and show them in Fig.~\ref{fig:profiling}(a). They showcase the non-linearity between throughput vs. message sizes we take into consideration in Sec.~\ref{design:slo-management}. 
As for $\frac{egressbw}{ingressbw}$, we find different accelerator types lead to different potential contention cases. For example, SHA-3-512 is much more likely to interfere with its ingress path (i.e., DMA reads) rather than on its egress path (i.e., DMA writes), which always has small messages as outputs. For another example, allocating $X$ Gbps PCIe bandwidth is not sufficient to feed data into a compression accelerator where accelerator SLO=$X$Gbps; the required minimal PCIe bandwidth depends on the compression ratio determined by the algorithm and implementation.

\myparagraph{Dynamism}
The \design{} mechanism allows re-configuring traffic shaping parameters. Our experiments show that the reconfiguration takes 10$\mu$s due to several rounds of PCIe transactions. This is quick enough to fit traffic pattern changes.

\input{tables/rate-limiter-comparison}

\subsubsection{Multi-flow Studies}\label{eva:multi-flow} 
We first study \myparagraph{Scalability} 
The \design{} interface requires only 0.97\% of the total ALMs (adaptive logic module) of our FPGA for each flow, including queues and the rate limiter. In addition, per-flow CPU overhead only takes 0.05 cores, thanks to hypervisor-bypassed architecture and traffic shaping offloading. Fig.~\ref{fig:profiling}(b) shows the overall throughput from 1 flow to 16 flows. Due to low per-flow overhead, we can achieve near-full throughput with 1 flow. Further optimizations to drivers and the accelerator interface are left for future work. 

When scaling to $N$ flows, the hardware overhead grows as \verb|O(N)|. Therefore, the \design{} interface scaling to tens of flows has low hardware overhead. Further queue sharing across flows can be explored in the future. 

When scaling the throughput of accelerators, per-flow hardware overhead does not change (i.e., \verb|O(1)|). The maximal throughput \design{} can support is when all PCIe lanes are fully utilized. In other words, \design{} interface is not the bottleneck of performance scaling.

\myparagraph{How the control plane makes decisions based on learned characteristics}
Fig.~\ref{fig:profiling}(c) shows the characterization results with combined factors considered, e.g., traffic pattern, PCIe bi-direction, flow numbers, and accelerator heterogeneity. Given such results, the control plane classifies that VM1 with 16 1KB flows on the NIC RX path plus VM2 with 4 4KB flows can have 50\%/50\% fair allocation (i.e., y-axis equal to 1). This traffic pattern combination can be classified as either \verb|SLO|-\verb|Friendly| or \verb|SLO|-\verb|Violating|, given what SLOs the two VM2 require. For example, if both VMs require $\frac{1}{2}$ of accelerator bandwidth, this combination is tagged as friendly; otherwise, it is tagged as \verb|SLO|-\verb|Violating|. This classification is re-run every time a new flow is registered. 

\myparagraph{Use case1: Streaming large messages}
In Fig.~\ref{fig:large-msg}, VM1 has one flow, sending 4KB accelerator I/Os. VM2 has one flow, sending message sizes ranging from 1KB to 512KB. Both flows are bi-directional, invoking one accelerator through function call mode. \design{} can precisely allocate half the throughput to each VM in all cases. This is achieved by the control plane pacing the right rate for different traffic patterns. In the Host$_{no\_TS}$ baseline, VM1 suffers from accelerator throughput loss of 36-67\% compared to what VM1 should have been allocated when VM2's messages are larger than 4KB. 
This is because VM2 steals more accelerator throughput by congesting PCIe and on-NIC buffer with larger messages. Similarly, VM1 steals VM2's accelerator throughput by 60\% when VM2 sends 1KB messages. 

\begin{figure}
  \centering
  \includegraphics[width=0.99\columnwidth]{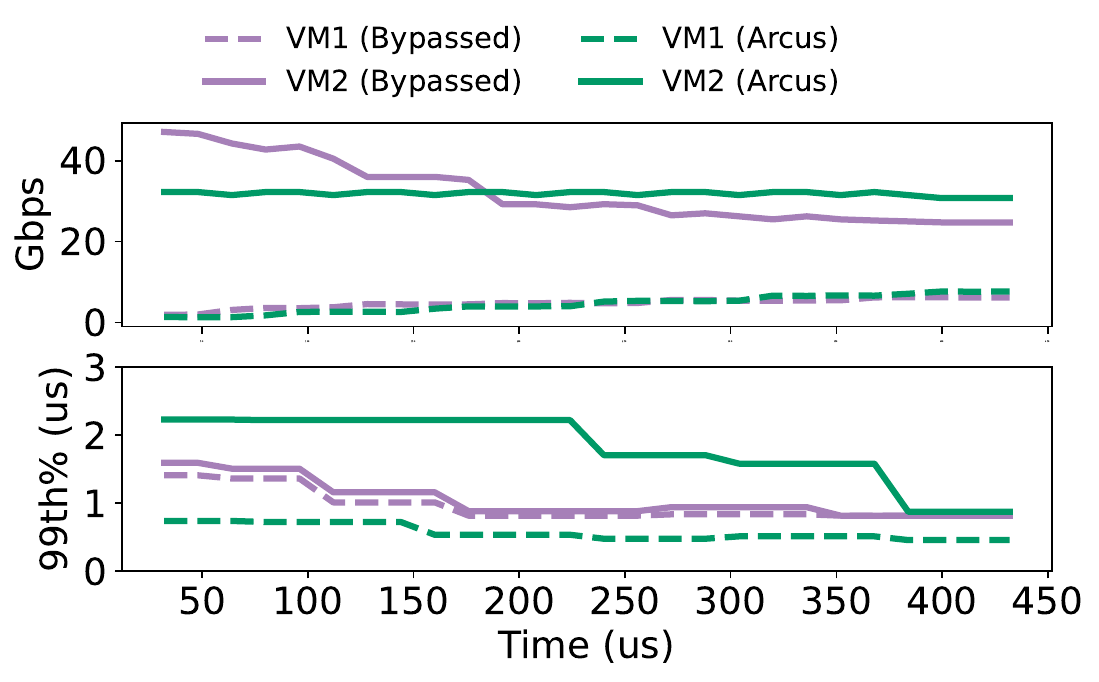}
  \caption{Use case2, SLO(64B)=1$\mu$s, SLO(1500B)=32Gbps. 
  }
     \label{fig:tiny-msg}
\end{figure}

\myparagraph{Use case2: Bursty tiny messages}
We co-locate two VMs, containing 64B and 1500B input data sizes, respectively. VM1 has one 64B flow, and VM2 has one 1500B flow. Both VMs run on the NIC RX path. VM1 is latency critical, whose 99th\% latency is restricted within 1$\mu$s~\cite{sutherland2020nebula,kalia2019erpc,ibanez2021nanopu}. As shown in Fig.~\ref{fig:tiny-msg}, the baseline Bypassed system fails to offer predictability. First, for VM1, although \design{} and Bypass both offer similar throughput, latency numbers are drastically different. \design{} guarantees VM1 has 0.5$\mu$s average latency, and 99th\% latency is no greater than 0.74$\mu$s. \design{} reduces 99th\% latency by up to 1.9$\times$, saving all SLO violations in the baseline. Second, \design{} guarantees stable throughput of VM2' 1500B stream. \design{} trades VM2's 99th\% latency for VM2's stable throughput and VM1's low tail latency. The sacrifice is acceptable for MTU-sized message streams which are throughput sensitive rather than latency sensitive~\cite{zhang2022justitia}. Before 200$\mu$s, \design{} avoids VM2 overloading the system, while Bypassed system fails to achieve, whose VM2 throughput is larger than 32Gbps. This overloading also causes an unnecessary 99th\% latency increase of VM1. With time going by, starting from 230th $\mu$s, long-tailed 1500B gets more stable, thus \design{} achieves similar 99th\% latency as Bypassed system does, after 380th $\mu$s.

%% file: tables/rate-limiter-comparison.tex
\begin{table}
  \centering\footnotesize
  \caption{VM1 (4KB streams) throughput deviation to the rate limit target, sharing an accelerator with VM2 (4KB stream). SLO$_{VM1}$:SLO$_{VM2}$=300K IOPS:200K IOPS. 
  }
  \begin{tabular}{ccccc}
  \toprule
                       & 25th\%               & 50th\%               & 75th\%               & 99th\%                \\
  \midrule
  Host$_{TS\_reflex}$   & \textcolor{purple}{-11.7\%}              & \textcolor{purple}{-7.3\%}               & -2.7\%               & \textcolor{purple}{+8.7\%}                \\
  Host$_{TS\_firecraker}$               & \textcolor{purple}{-6.7\%}               & +2.7\%               & \textcolor{purple}{+4.7\%}               & \textcolor{purple}{+24.3\%}               \\
  \design{}         & -0.7\%               & -0.3\%               & +0.3\%               & +0.7\%                \\
  \bottomrule 
\end{tabular}
\label{table:rate-limit-comparison}
\end{table}

%% file: experiment-sections/ev-ete.tex
\begin{figure}
    \centering
    \includegraphics[width=0.99\columnwidth]{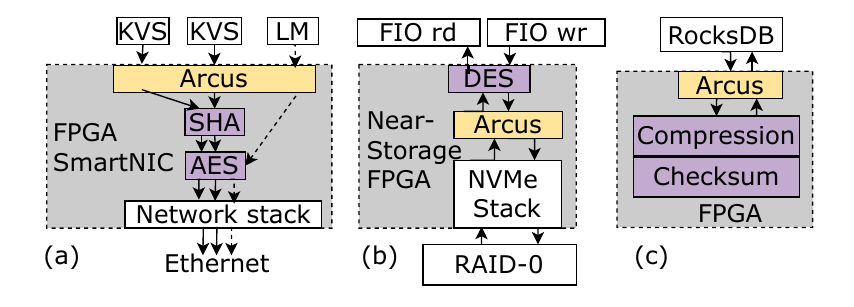}
    \caption{Prototypes built for end-to-end evaluation. 
    }
    \label{fig:config-prototype}
\end{figure}

\begin{figure}
\centering 
\subfigure[FPGA SmartNIC Mops@SLO]{
    \includegraphics[width=0.6\columnwidth]{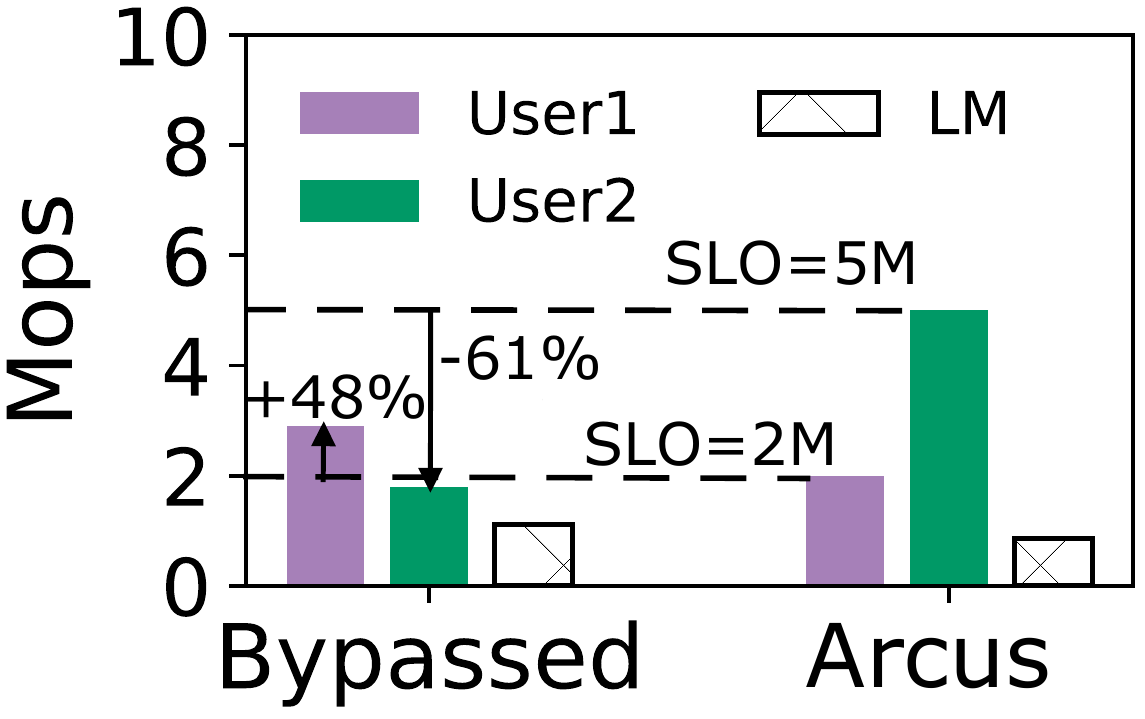}}
\subfigure[Near-SSD FPGA KIOPS@SLO]{
    \includegraphics[width=0.6\columnwidth]{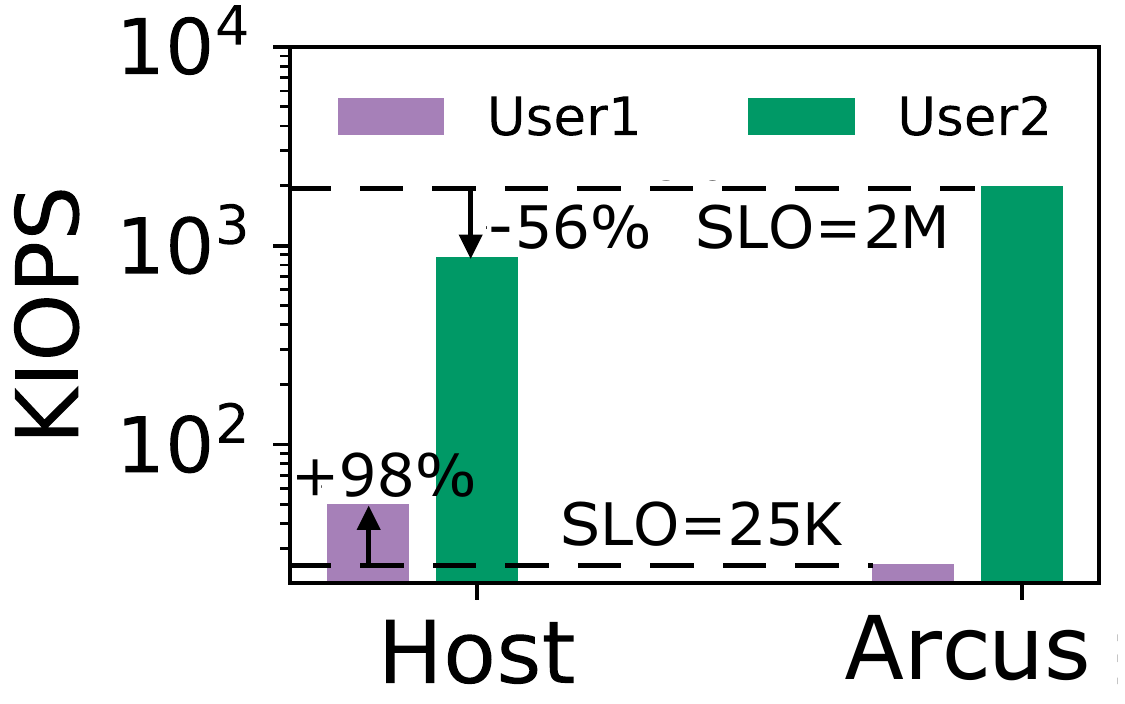}
}
\caption{Results for network and storage applications. 
}
\label{fig:ete}
\end{figure}

\myparagraph{Inline NIC mode: MICA and live migration}
We built an FPGA-based SmartNIC prototype (Fig.~\ref{fig:config-prototype}(a)) for inline acceleration. The FPGA has two 50Gbps Ethernet ports. As a bump-in-the-wire architecture, the FPGA sits between the CPU and the network, handling network packets based on LTL reliable protocol~\cite{caulfield2016ltl}. This experiment represents inter-VM intra-path contention. Two users run low-latency MICA~\cite{sutherland2020nebula,daglis2019rpcvalet,lim2014mica}, each with 50/50 GET/SET. The value sizes are 64B and 256B for user1 and user2, respectively. Two users share two accelerators,  SHA1-HMAC and AES-128-CBC, required by secure network applications~\cite{son2017protego,lin2020panic}. In addition, another live migration (LM) is co-running~\cite{bindschaedler2020hailstorm,annamalai2018akkio,kulkarni2017rocksteady}, contending for the AES accelerator with two MICA users. The LM job sends MTU-sized large messages to the Ethernet, i.e., 1500B. In Fig.~\ref{fig:ete}(a), we show maximal MOps at which 99th\% latency smaller than 10$\times$ average latency, used by prior work~\cite{tail-at-scale,sutherland2020nebula,zhao2022altocumulus}. We compare \design{} with a hypervisor-bypassed system with PANIC~\cite{lin2020panic} interface, which is well-optimized for tiny network messages.

Fig.~\ref{fig:ete}(a) shows \design{}  satisfies SLO accurately for two MICA users. The baseline over-provisions 48\% throughput for user1, while user2 loses 61\%. This is due to the traffic pattern mixture in the PANIC interface and PCIe. In addition, although we prioritize MICA traffic over LM in the baseline, LM's large message stream still overloads the accelerator sub-system, interfering with latency-sensitive MICA users.

\textit{\textbf{Key takeaways}}:
(1) Our approach can augment SLO management for existing SmartNIC accelerators~\cite{liu2019ipipe,grant2020fairnic,lin2020panic,eran2019nica}. (2) \design{} allows accelerators to be safely shared between users and the provider. Remaining throughput can be harvested by background tasks such as LM. 

\myparagraph{Inline (P2P) mode: storage reads and writes}
We set another prototype based on Fig.~\ref{fig:config-prototype}(b), where two users run reads and writes, respectively, based on FIO~\cite{fio}. This setting represents inter-VM intra-path contention on two workloads that are DMA read- and DMA write-intensive. The two users share a RAID-0 storage system composed of four Samsung 983 DCT SSDs. We build an NVMe stack based on FVM~\cite{kwon2020fvm}. 

The SLO we set is SLO2$_{read}$ = 2 Million IOPS, SLO2$_{write}$ = 25 Kilo IOPS (KIOPS). Reads are 1KB random reads; writes are 4KB sequential writes. In Fig.~\ref{fig:ete}(b), we compare maximal IOPS at which 99th\% latency $<$2ms across two users. 

\design{} realizes SLO numbers for 99th\% latency $<$ 2ms. The baseline over-provisions write throughput up to 50 KIOPS, while read throughput falls short at only 44\% of the required SLO2$_{read}$. In addition, the overall RAID-0 throughput is only 930 KIOPS in the baseline, degraded by 2.2$\times$. 

\textit{\textbf{Key takeaways}}:
(1) \design{} can re-shape injection rates difference of 4.4$\times$ between two users due to PCIe contention to be what SLO requires. (2) For storage SLOs, it's even more vital to keep reads/writes throughput precisely. Otherwise the overall throughput will be significantly affected. The root cause is internal read-write interference in SSD sub-systems~\cite{min2021gimbal}. This makes IOPS even more unpredictable.

\myparagraph{Function call mode: RocksDB compression and checksum}
We test RocksDB instances on an 8-core VM. In this experiment, \design{}-enabled system is built based on Fig.~\ref{fig:config-prototype}(c), where RocksDB can offload checksum and compression onto an FPGA (i.e., function call mode). The baseline runs RocksDB based on ext4. 
As Table~\ref{table:ete-rocksdb} shows, ext4 delivers 161.7 MB/s, which is 1.43$\times$ lower than \design{}-enabled system. In addition, due to accelerator offloads, \design{} saves 58.9\% VM cores. \design{} software only uses 17.5\% of a core.

\textit{\textbf{Key takeaways}}:
(1) \design{} enables accelerator offloads with low CPU overheads. (2) More cores can be used for applications, rather than checksum and compression.

\input{tables/eva-rocksdb}

%% file: tables/eva-rocksdb.tex
\begin{table}
\centering\footnotesize
\caption{RocksDB evaluation on a 8-core VM. 
}
\vspace{-1em}
\begin{tabular}{cccc}
\toprule
    & ext4  & \design{}-enabled & Benefits        \\
\midrule
    RocksDB Thr. (MB/s) & 161.7 & 231.2         & 1.43$\times$       \\
    \# CPU cores used              & 5.23  & 2.15~         & 58.9\% savings \\
\bottomrule
\end{tabular}
\label{table:ete-rocksdb}
\end{table}

%% file: other-sections/discussions.tex
\myparagraph{Enabling accelerator SLO policies}
\design{} enables user-facing SLO policies like other compute resources~\cite{carvalho2014longtermSLO,azure-managed-burst}. 

\begin{itemize}
    \item \textit{Reserved}. Give up to a certain throughput  with  availability of almost 100\% for long-term commitments. 
    \item \textit{On-demand}. Provide a service with high availability (e.g., 99\%), and with short-term commitments to a certain performance level, involving frequent (de)allocations. 
    \item \textit{Managed burst}. For example, allow bursting throughput from $X$ to 10$X$ for 30 minutes per day~\cite{azure-managed-burst}. 
    \item \textit{Opportunistic}. No SLO performance guarantee over a time window, but useful to improve utilization. 
\end{itemize}

\myparagraph{\design{} deployment}
We emphasize that \design{} interface should be provider-managed. It can be implemented in (1) discrete accelerator cards like FPGAs or ASICs, (2) NIC/GPU/DPU/IPU, to achieve precise SLO management. However, many devices are vendor-specific third-party designs cannot be controlled by providers. SmartNICs have been customized for different management goals~\cite{aws-nitro,azuresmartnic,zhao2024smartniclivemigration,kim2021linefs,kwon2020fvm,caulfield2016ltl,li2020leapio,wu2024tomur}. Our rationale is that building \design{} interface on SmartNICs is a viable and practical option because they are provider-managed, vendor-agnostic and customizeable.
First, SmartNICs are already in many clouds~\cite{aws-nitro,azuresmartnic}, adding no extra capital cost. Second, SmartNICs offer natural network accessibility, opening up design spaces for allowing any VM to use any accelerator. Third, cloud providers already customize their own SmartNICs with critical features in public clouds~\cite{aws-nitro,azuresmartnic,zhao2024smartniclivemigration}. Successful production-scale SmartNICs~\cite{aws-nitro,azuresmartnic} provide such feasibility. For example, virtual networking offloaded on FPGA-based SmartNICs significantly improves performance predictability. 

\myparagraph{Why a hardware interface rather than SoC cores}
Using on-device SoC cores like ARM or RISC-V cores~\cite{liu2019ipipe,liu2019e3,khalilov2023osmosis,li2020leapio} still suffers from imprecise accelerator provisioning. This might be acceptable for good fairness or isolation with predictable average performance, but not suitable for SLO-oriented systems. \design{} leverages a new piece of hardware to mediate accelerator traffic. It was partially inspired by shim layers previously proposed~\cite{zhang2022justitia,kumar2019picnic}, but they run in software and do not work for accelerators. Hardware also reduces host contention, improves predictability~\cite{azuresmartnic}. The limitation of hardware's inflexibility and high overheads is mitigated by our programmable and low-cost designs, evaluated in Sec.~\ref{eva:breakdown}. 

\myparagraph{Accelerator management}
We list related work of \design{} in Table~\ref{table:prior-art}. 
Some of them~\cite{zhang2022justitia,kumar2019picnic,khalilov2023osmosis} are not designed for accelerators, though they have multi-tenancy and isolation supports. While hypervisor-managed works can provide end-to-end traffic probing~\cite{zhang2022justitia,kumar2019picnic,ji2023yama}, they need tenant information a priori, which is not suitable for the cloud scenario. Hypervisor-bypassed systems use reactive scheduling schemes for compute and memory resource management~\cite{grant2020fairnic,khalilov2023osmosis,lin2020panic,intel-quickassist,kwon2021flexcsv,chen2024dpa}, while \design{} can enable hypervisor-like SLO manageability with proactive traffic shaping. No prior work takes accelerator heterogeneity into account. Some other work not listed in Table~\ref{table:prior-art} also discusses the PCIe contention~\cite{zhao2024accelerator-as-a-service,hou2024rpciebench,neugebauer2018understandingpcie,tian2021cloudfpga-pcie-contention}, but they do not aim to guarantee SLOs and support heterogeneous accelerators. 


\myparagraph{SLO: throughput vs.~latency}
Generally, the SLO can be defined as tail latency, throughput, drop rate, etc., in different contexts~\cite{tail-at-scale,kumar2019picnic}. Our work mainly targeting throughput guarantee as the SLO target is reasonable. This is because many users pursue datacenter tax offload primarily for benefits such as saving CPU cycles and throughput improvement. Such scenario that \design{} supports is orthogonal with other scenarios when VM users want ultra low-latency accelerator services. In those cases, hosting such VMs on CPUs with sufficient on-chip accelerator resources will be helpful~\cite{sapphirerapids-accelerator-profiling,sriraman2020accelerometer,karandikar2023cdpu}. 
Meanwhile \design{} does significantly shorten and stabilize latency like prior work justifies~\cite{azuresmartnic}; our evaluation results in Fig.~\ref{fig:tiny-msg} corroborates this claim.

%% file: other-sections/related-work.tex
\myparagraph{Traffic shaping}
The technique is useful for handling datacenter traffic with varied-size packets and unpredictable burstness, across distributed hosts that do not necessarily have sufficient network QoS supports in between~\cite{radhakrishnan2014senic,ballani2011oktopus,angel2014pulsar,kumar2019picnic}. For example, cloud providers allow users to request rate limiting service~\cite{aws-ebs-ratelimit,azure-ratelimit,google-cloud-ratelimit,nginx-ratelimit}. We are the first to leverage traffic shaping for accelerators in VM environments and SLO-oriented scenarios. We do not directly borrow the technique, but construct a holistic protocol that incorporates heterogeneity and communication resource characteristics.

\input{tables/smartnic-acceleration-opportunity}

\myparagraph{Accelerator prevalence}
Table~\ref{table:smartnic-acceleration-opportunities} lists various compute-intensive tasks required by either users or cloud providers, or both.
The table shows their CPU utilization and their potential for offloading to an I/O accelerator.
For instance, it is possible to offload 3-15\% of CPU cycles by using (de)compression accelerators.
Also, the IPSec accelerator offering 32Gbps throughput dramatically outperforms the cost efficiency of CPUs, which can only deliver 10Gbps while consuming as many as 8 Xeon
cores.

\myparagraph{Accelerator management}
We list related work of \design{} in Table~\ref{table:prior-art}. Some of them~\cite{zhang2022justitia,kumar2019picnic,khalilov2023osmosis} are not designed for accelerators, though they have multi-tenancy and isolation supports. While hypervisor-managed works can provide end-to-end traffic probing~\cite{zhang2022justitia,kumar2019picnic,ji2023yama}, they need tenant information a priori, which is not suitable for the cloud scenario. Hypervisor-bypassed systems use reactive scheduling schemes for compute and memory resource management~\cite{grant2020fairnic,khalilov2023osmosis,lin2020panic,intel-quickassist,kwon2021flexcsv,chen2024dpa}, while \design{} can enable hypervisor-like SLO manageability with proactive traffic shaping. No prior work takes accelerator heterogeneity into account. Some other work not listed in Table~\ref{table:prior-art} also discusses the PCIe contention~\cite{zhao2024accelerator-as-a-service,hou2024rpciebench,neugebauer2018understandingpcie,tian2021cloudfpga-pcie-contention}, but they do not aim to guarantee SLOs and support heterogeneous accelerators.

Another line of research diversifies the path supported by the management layer~\cite{chen2024dpa,eran2022flexdriver,eran2019nica,tork2020lynx,lim2024beehive,wang2022fpganic,pismenny2021autonomous,sadok2023enso}. However, they do not support multi-tenancy and SLO management.

%% file: tables/smartnic-acceleration-opportunity.tex
\begin{table}
    \centering
    \footnotesize
    \caption{Acceleration chances on accelerators.}
    \begin{tabular}{ccc} 
    \toprule
    \begin{tabular}[c]{@{}c@{}}\textbf{Some Tasks at}\\\textbf{Google~\cite{karandikar2021grpcacceleration,karandikar2023cdpu}}\end{tabular} & \begin{tabular}[c]{@{}c@{}}\textbf{CPU }\\\textbf{Usage}\end{tabular} & \begin{tabular}[c]{@{}c@{}}\textbf{Acceleration }\\\textbf{Opportunity}\end{tabular} \\

    \midrule
    \textbf{(De)compression}                                                         & 2.9\%                                                                      &  (De)compressor~\cite{dpu-3}                                        \\ 
    \textbf{(De)serialization}                                                                 & 3.45\%                                                                       &     NIC Scatter-gather~\cite{raghavan2021breakfast}                                     \\ 
    \midrule
    \midrule
    \begin{tabular}[c]{@{}c@{}}\textbf{Some Tasks at}\\\textbf{Meta~\cite{sriraman2020accelerometer}}\end{tabular} & \begin{tabular}[c]{@{}c@{}}\textbf{CPU }\\\textbf{Usage}\end{tabular} & \begin{tabular}[c]{@{}c@{}}\textbf{Acceleration }\\\textbf{Opportunity}\end{tabular} \\
    
    \midrule
    \textbf{(De)compression}                                                         & 3-15\%                                                                      & ZIP~\cite{intel-quickassist} or GPUs~\cite{gpu-nvcomp-compression}                                         \\ 
    \textbf{Hashing}                                                                 & 1-4\%                                                                       & SHA~\cite{dpu-3}                                        \\ 
    \textbf{I/O Send/Recv.}                                                          & 4-52\%                                                                      & Special logic                          \\ 
    \textbf{(De)serialization}                                                       & 4-13\%                                                                      & Serializer~\cite{wolnikowski2021zerializer}                                  \\ 
    \textbf{Encryption}                                                              & 1-6\%                                                                       & AES-256~\cite{dpu-3}                                     \\ 
    \textbf{Ranking}                                                                 & 33-58\%                                                                     & 3DES/NN~\cite{liquidio,dpu-3}                      \\ 
    \midrule
    \midrule
    \textbf{Other Tasks}                                                             & \textbf{\textbf{CPU Usage}}                                                 & \textbf{\textbf{Acc. Opportunity}}  \\ 
    \midrule
    \textbf{IP Security~\cite{son2017protego}
    }                                                             & 8 cores (10Gbps)                                                         & IPSec (32Gbps)
    \\ 
    \textbf{Data Copy~\cite{kwon2020fvm}}                                                           & 2 cores (18Gbps)                                                         & Special logic (18Gbps)                   \\
    \bottomrule
    \end{tabular}
    \label{table:smartnic-acceleration-opportunities}
\end{table}

%% file: tables/prior-art.tex
\begin{table}
\centering\footnotesize
\caption{Comparing \design{} with other related work.  
}
\vspace{-1em}
\begin{tabular}{ccccc}
\toprule
                       System
                       & \begin{tabular}[c]{@{}c@{}}SLO \\Management\end{tabular}                                                                                                                                     & \begin{tabular}[c]{@{}c@{}}Acc. \\Profiling\end{tabular}                             & \begin{tabular}[c]{@{}c@{}}Acc. \\Sched.\end{tabular}                                                                                                                                                                                 & \begin{tabular}[c]{@{}c@{}}Comm. \\Conten.\end{tabular}                                     \\
\midrule
Justitia~\cite{zhang2022justitia}               & \multirow{3}{*}{\begin{tabular}[c]{@{}c@{}}Hypervisor- \\managed\end{tabular}} & No & \multicolumn{2}{c}{End-to-end proactive}\\
PicNIC~\cite{kumar2019picnic}                 &  & No & \multicolumn{2}{c}{traffic probe \& mgmt,} \\
YAMA~\cite{ji2023yama}                   &   & No & \multicolumn{2}{c}{requiring tenant info.}  \\
\midrule
QAT~\cite{intel-quickassist} & \multirow{2}{*}{\begin{tabular}[c]{@{}c@{}}SR-IOV, \\no SLO policies \end{tabular}} & No & Reactive & No  \\
FlexCSV~\cite{kwon2021flexcsv}  & & No & Reactive & No  \\
\midrule
FairNIC~\cite{grant2020fairnic} & \multirow{2}{*}{\begin{tabular}[c]{@{}c@{}}Device SW, \\ no SLO policies\end{tabular}} & No & Reactive & No \\
OSMOSIS~\cite{khalilov2023osmosis}                &  & No & Reactive & Async.            \\
\midrule
PANIC~\cite{lin2020panic}                 & \begin{tabular}[c]{@{}c@{}}Device HW, \\ dynamic policies\end{tabular} & No & Reactive & No \\
\midrule
\design{}                  & \begin{tabular}[c]{@{}c@{}}Device HW, \\accurate \&  \\dynamic policies \end{tabular} & 
\multicolumn{3}{c}{\begin{tabular}[c]{@{}c@{}}Profiling acc. \& comm. conten.,\\decoupled proactive traffic mgnt,\\ lightweight \& central interface\end{tabular}}  \\  
\bottomrule
\end{tabular}
\label{table:prior-art}
\end{table}

%% file: main.bbl
\begin{thebibliography}{10}

\bibitem{intel-quickassist}
Intel quickassist technology.
\newblock \url{https://01.org/ intel-quickassist-technology}, 2020.
\newblock Accessed: 2024-8-17.

\bibitem{adatapcie5}
Adata pcie gen5x 4 ssds (project nighthawk).
\newblock \url{https://www.adata.com/en/news/960}, 2021.
\newblock Accessed: 2024-8-19.

\bibitem{dpu-3}
Nvidia bluefield-3 dpu programmable data center infrastructure on-a-chip.
\newblock \url{https://www.nvidia.com/content/dam/en-zz/Solutions/Data-Center/documents/datasheet-nvidia-bluefield-3-dpu.pdf}, 2021.
\newblock Accessed: 2024-7-11.

\bibitem{samsungpcie5}
Samsung high-performance pcie 5.0 ssd for enterprise servers.
\newblock \url{https://news.samsung.com/global/samsung-develops-high-performance-pcie-5-0-ssd-for-enterprise-servers}, 2021.
\newblock Accessed: 2024-8-17.

\bibitem{liquidio}
Marvell liquidio smartnic.
\newblock \url{https://www.marvell.com/products/ethernet-adapters-and-controllers/liquidio-smart-nics.html}, 2022.
\newblock Accessed: 2022-7-8.

\bibitem{google-cloud-ratelimit}
Google cloud rate limiting.
\newblock \url{https://cloud.google.com/architecture/infra-reliability-guide/traffic-load}, 2023.
\newblock Accessed: 2024-8-17.

\bibitem{azure-managed-burst}
Managed disk bursting for azure vms.
\newblock \url{https://learn.microsoft.com/en-us/azure/virtual-machines/disk-bursting}, 2023.
\newblock Accessed: 2024-8-17.

\bibitem{aws-f1}
Amazon ec2 f1 instances.
\newblock \url{https://aws.amazon.com/ec2/instance-types/f1/}, 2024.
\newblock Accessed: 2024-7-11.

\bibitem{aws-ebs-ratelimit}
Amazon elastic block store: Easy to use, high perfromance block storage at any scale.
\newblock \url{https://aws.amazon.com/ebs/}, 2024.
\newblock Accessed: 2024-8-17.

\bibitem{amd-xilinx-smartnic}
Amd xilinx alveo sn1000 smartnic.
\newblock \url{https://www.xilinx.com/applications/data-center/network-acceleration/alveo-sn1000.html}, 2024.
\newblock Accessed: 2022-12-30.

\bibitem{aws-nitro}
Aws nitro system.
\newblock \url{https://aws.amazon.com/ec2/nitro/}, 2024.
\newblock Accessed: 2024-8-17.

\bibitem{azure-ratelimit}
Azure limit call rate by subscription.
\newblock \url{https://learn.microsoft.com/en-us/azure/api-management/rate-limit-policy}, 2024.
\newblock Accessed: 2024-8-17.

\bibitem{ethernet-roadmap}
Ethernet roadmap 2024.
\newblock \url{https://ethernetalliance.org/technology/ethernet-roadmap/}, 2024.
\newblock Accessed: 2024-8-19.

\bibitem{fio}
Flexible i/o tester.
\newblock \url{https://github.com/axboe/fio}, 2024.
\newblock Accessed: 2024-8-17.

\bibitem{intel-ipu}
Intel infrastructure processing unit (intel ipu) and smartnics.
\newblock \url{https://www.intel.com/content/www/us/en/products/details/network-io/ipu.html}, 2024.
\newblock Accessed: 2024-7-11.

\bibitem{gpu-nvcomp-compression}
Nvidia nvcomp library documentation.
\newblock \url{https://docs.nvidia.com/cuda/nvcomp/index.html}, 2024.
\newblock Accessed: 2024-9-30.

\bibitem{token-bucket}
Token bucket algorithm.
\newblock \url{https://en.wikipedia.org/wiki/Token_bucket}, 2024.
\newblock Accessed: 2024-8-17.

\bibitem{agache2020firecracker}
Alexandru Agache, Marc Brooker, Alexandra Iordache, Anthony Liguori, Rolf Neugebauer, Phil Piwonka, and Diana-Maria Popa.
\newblock Firecracker: Lightweight virtualization for serverless applications.
\newblock In {\em Proceedings of the 17th USENIX Symposium on Networked Systems Design and Implementation (NSDI)}, pages 419--434, 2020.

\bibitem{sapphirerapids-accelerator-profiling}
Dave Altavilla and Marco Chiappetta.
\newblock Live intel 4th gen xeon benchmarks: Sapphire rapids accelerators revealed.
\newblock \url{https://hothardware.com/news/intel-4th-gen-xeon-sapphire-rapids-accelerator-benchmarks}, 2022.
\newblock Accessed: 2024-8-19.

\bibitem{angel2014pulsar}
Sebastian Angel, Hitesh Ballani, Thomas Karagiannis, Greg O'Shea, and Eno Thereska.
\newblock End-to-end performance isolation through virtual datacenters.
\newblock In {\em Proceedings of the 11th USENIX Symposium on Operating Systems Design and Implementation (OSDI)}, pages 233--248, 2014.

\bibitem{annamalai2018akkio}
Muthukaruppan Annamalai, Kaushik Ravichandran, Harish Srinivas, Igor Zinkovsky, Luning Pan, Tony Savor, David Nagle, and Michael Stumm.
\newblock Sharding the shards: managing datastore locality at scale with akkio.
\newblock In {\em Proceedings of the 13th USENIX Symposium on Operating System Design and Implementation (OSDI)}, pages 445--460, 2018.

\bibitem{ballani2011oktopus}
Hitesh Ballani, Paolo Costa, Thomas Karagiannis, and Ant Rowstron.
\newblock Towards predictable datacenter networks.
\newblock In {\em Proceedings of the ACM SIGCOMM Conference}, pages 242--253, 2011.

\bibitem{aws-aqua}
Jeff Barr.
\newblock Aqua (advanced query accelerator) – a speed boost for your amazon redshift queries.
\newblock \url{https://aws.amazon.com/blogs/aws/new-aqua-advanced-query-accelerator-for-amazon-redshift/}, 2021.
\newblock Accessed: 2024-8-17.

\bibitem{bindschaedler2020hailstorm}
Laurent Bindschaedler, Ashvin Goel, and Willy Zwaenepoel.
\newblock Hailstorm: Disaggregated compute and storage for distributed lsm-based databases.
\newblock In {\em Proceedings of the 25th International Conference on Architectural Support for Programming Languages and Operating Systems (ASPLOS)}, pages 301--316, 2020.

\bibitem{carvalho2014longtermSLO}
Marcus Carvalho, Walfredo Cirne, Francisco Brasileiro, and John Wilkes.
\newblock Long-term slos for reclaimed cloud computing resources.
\newblock In {\em Proceedings of the 5th ACM Symposium on Cloud Computing (SoCC)}, pages 1--13, 2014.

\bibitem{caulfield2016ltl}
Adrian~M. Caulfield, Eric~S. Chung, Andrew Putnam, Hari Angepat, Jeremy Fowers, Michael Haselman, Stephen Heil, Matt Humphrey, Puneet Kaur, Joo-Young Kim, Daniel Lo, Todd Massengill, Kalin Ovtcharov, Michael Papamichael, Lisa Woods, Sitaram Lanka, Derek Chiou, and Doug Burger.
\newblock A cloud-scale acceleration architecture.
\newblock In {\em Proceedings of the 49th IEEE/ACM International Symposium on Microarchitecture (MICRO)}, pages 1--13, 2016.

\bibitem{chen2024dpa}
Xuzheng Chen, Jie Zhang, Ting Fu, Yifan Shen, Shu Ma, Kun Qian, Lingjun Zhu, Chao Shi, Yin Zhang, Ming Liu, and Zeke Wang.
\newblock Demystifying datapath accelerator enhanced off-path smartnic.
\newblock {\em arXiv preprint arXiv:2402.03041}, 2024.

\bibitem{azure-corsica-compression-asic}
Derek Chiou, Eric Chung, and Carrie. Susan.
\newblock (cloud) acceleration at microsoft.
\newblock \url{https://old.hotchips.org/hc31/HC31_T2_Microsoft_CarrieChiouChung.pdf}, 2019.
\newblock Accessed: 2024-8-17.

\bibitem{chung2018brainwave}
Eric Chung, Jeremy Fowers, Kalin Ovtcharov, Michael Papamichael, Adrian Caulfield, Todd Massengill, Ming Liu, Daniel Lo, Shlomi Alkalay, Michael Haselman, Maleen Abeydeera, Logan Adams, Hari Angepat, Christian Boehn, Derek Chiou, Oren Firestein, Alessandro Forin, Kang~Su Gatlin, Mahdi Ghandi, Stephen Heil, Kyle Holohan, Ahmad~El Husseini, Tamas Juhasz, Kara Kagi, Ratna~K. Kovvuri, Sitaram Lanka, Friedel van Megen, Dima Mukhortov, Prerak Patel, Brandon Perez, Amanda~Grace Rapsang, Steven~K. Reinhardt, Bita~Darvish Rouhani, Adam Sapek, Raja Seera, Sangeetha Shekar, Balaji Sridharan, Gabriel Weisz, Lisa Woods, Phillip~Yi Xiao, Dan Zhang, Ritchie Zhao, and Doug Burger.
\newblock Serving dnns in real time at datacenter scale with project brainwave.
\newblock {\em IEEE Micro}, 38(2):8--20, 2018.

\bibitem{daglis2019rpcvalet}
Alexandros Daglis, Mark Sutherland, and Babak Falsafi.
\newblock Rpcvalet: Ni-driven tail-aware balancing of $\mu$s-scale rpcs.
\newblock In {\em Proceedings of the 24th International Conference on Architectural Support for Programming Languages and Operating Systems (ASPLOS)}, pages 35--48, 2019.

\bibitem{tail-at-scale}
Jeffrey Dean and Luiz~Andr{\'e} Barroso.
\newblock The tail at scale.
\newblock {\em Communications of the ACM}, 56(2):74--80, 2013.

\bibitem{eran2022flexdriver}
Haggai Eran, Maxim Fudim, Gabi Malka, Gal Shalom, Noam Cohen, Amit Hermony, Dotan Levi, Liran Liss, and Mark Silberstein.
\newblock Flexdriver: a network driver for your accelerator.
\newblock In {\em Proceedings of the 27th ACM International Conference on Architectural Support for Programming Languages and Operating Systems (ASPLOS)}, pages 1115--1129, 2022.

\bibitem{eran2019nica}
Haggai Eran, Lior Zeno, Maroun Tork, Gabi Malka, and Mark Silberstein.
\newblock Nica: An infrastructure for inline acceleration of network applications.
\newblock In {\em Proceedings of the 2019 USENIX Annual Technical Conference (ATC)}, pages 345--362, 2019.

\bibitem{esmaeilzadeh2011darksilicon}
Hadi Esmaeilzadeh, Emily Blem, Renee~St Amant, Karthikeyan Sankaralingam, and Doug Burger.
\newblock Dark silicon and the end of multicore scaling.
\newblock In {\em Proceedings of the 38th ACM/IEEE International Symposium on Computer Architecture (ISCA)}, pages 365--376, 2011.

\bibitem{azuresmartnic}
Daniel Firestone, Andrew Putnam, Sambhrama Mundkur, Derek Chiou, Alireza Dabagh, Mike Andrewartha, Hari Angepat, Vivek Bhanu, Adrian Caulfield, Eric Chung, Harish~Kumar Chandrappa, Somesh Chaturmohta, Matt Humphrey, Jack Lavier, Norman Lam, Fengfen Liu, Kalin Ovtcharov, Jitu Padhye, Gautham Popuri, Shachar Raindel, Tejas Sapre, Mark Shaw, Gabriel Silva, Madhan Sivakumar, Nisheeth Srivastava, Anshuman Verma, Qasim Zuhair, Deepak Bansal, Doug Burger, Kushagra Vaid, David~A. Maltz, and Albert Greenberg.
\newblock Azure accelerated networking: Smartnics in the public cloud.
\newblock In {\em Proceedings of the 15th USENIX Symposium on Networked Systems Design and Implementation (NSDI)}, pages 51--66, 2018.

\bibitem{galles2021pensando}
Michael Galles and Francis Matus.
\newblock Pensando distributed services architecture.
\newblock {\em IEEE Micro}, 41(2):43--49, 2021.

\bibitem{grant2020fairnic}
Stewart Grant, Anil Yelam, Maxwell Bland, and Alex~C. Snoeren.
\newblock Smartnic performance isolation with fairnic: Programmable networking for the cloud.
\newblock In {\em Proceedings of the ACM SIGCOMM Conference}, pages 681--693, 2020.

\bibitem{hou2024rpciebench}
Wentao Hou, Jie Zhang, Zeke Wang, and Ming Liu.
\newblock Understanding routable pcie performance for composable infrastructures.
\newblock In {\em Proceedings of the 21st USENIX Symposium on Networked Systems Design and Implementation (NSDI)}, pages 297--312, 2024.

\bibitem{ibanez2021nanopu}
Stephen Ibanez, Alex Mallery, Serhat Arslan, Theo Jepsen, Muhammad Shahbaz, Changhoon Kim, and Nick McKeown.
\newblock The nanopu: A nanosecond network stack for datacenters.
\newblock In {\em Proceedings of the 15th Symposium on Operating Systems Design and Implementation (OSDI)}, pages 239--256, 2021.

\bibitem{ji2023yama}
Tao Ji, Divyanshu Saxena, Brent~E. Stephens, and Aditya Akella.
\newblock Yama: Providing performance isolation for black-box offloads.
\newblock In {\em Proceedings of the 14th ACM Symposium on Cloud Computing (SoCC)}, pages 572--587, 2023.

\bibitem{kalia2019erpc}
Anuj Kalia, Michael Kaminsky, and David Andersen.
\newblock Datacenter rpcs can be general and fast.
\newblock In {\em Proceedings of the 16th USENIX Symposium on Networked Systems Design and Implementation (NSDI)}, pages 1--16, 2019.

\bibitem{kanev2015profiling}
Svilen Kanev, Juan~Pablo Darago, Kim Hazelwood, Parthasarathy Ranganathan, Tipp Moseley, Gu-Yeon Wei, and David Brooks.
\newblock Profiling a warehouse-scale computer.
\newblock In {\em Proceedings of the 42nd ACM/IEEE International Symposium on Computer Architecture (ISCA)}, pages 158--169, 2015.

\bibitem{karandikar2021grpcacceleration}
Sagar Karandikar, Chris Leary, Chris Kennelly, Jerry Zhao, Dinesh Parimi, Borivoje Nikolic, Krste Asanovic, and Parthasarathy Ranganathan.
\newblock A hardware accelerator for protocol buffers.
\newblock In {\em Proceedings of the 54th ACM/IEEE International Symposium on Microarchitecture (MICRO)}, pages 462--478, 2021.

\bibitem{karandikar2023cdpu}
Sagar Karandikar, Aniruddha~N. Udipi, Junsun Choi, Joonho Whangbo, Jerry Zhao, Svilen Kanev, Edwin Lim, Jyrki Alakuijala, Vrishab Madduri, Yakun~Sophia Shao, Borivoje Nikolic, Krste Asanovic, and Parthasarathy Ranganathan.
\newblock Cdpu: Co-designing compression and decompression processing units for hyperscale systems.
\newblock In {\em Proceedings of the 50th ACM/IEEE International Symposium on Computer Architecture (ISCA)}, pages 1--17, 2023.

\bibitem{khalilov2023osmosis}
Mikhail Khalilov, Marcin Chrapek, Siyuan Shen, Alessandro Vezzu, Thomas Benz, Salvatore Di~Girolamo, Timo Schneider, Daniele De~Sensi, Luca Benini, and Torsten Hoefler.
\newblock Osmosis: Enabling multi-tenancy in datacenter smartnics.
\newblock In {\em Proceedings of the 2024 USENIX Annual Technical Conference (ATC)}, pages 247--263, 2024.

\bibitem{kim2021linefs}
Jongyul Kim, Insu Jang, Waleed Reda, Jaeseong Im, Marco Canini, Dejan Kosti{\'c}, Youngjin Kwon, Simon Peter, and Emmett Witchel.
\newblock Linefs: Efficient smartnic offload of a distributed file system with pipeline parallelism.
\newblock In {\em Proceedings of the ACM SIGOPS 28th Symposium on Operating Systems Principles (SOSP)}, pages 756--771, 2021.

\bibitem{klimovic2017reflex}
Ana Klimovic, Heiner Litz, and Christos Kozyrakis.
\newblock Reflex: Remote flash $\approx$ local flash.
\newblock {\em Proceedings of the 22nd ACM International Conference on Architectural Support for Programming Languages and Operating Systems (ASPLOS)}, pages 345--359, 2017.

\bibitem{kulkarni2017rocksteady}
Chinmay Kulkarni, Aniraj Kesavan, Tian Zhang, Robert Ricci, and Ryan Stutsman.
\newblock Rocksteady: Fast migration for low-latency in-memory storage.
\newblock In {\em Proceedings of the 26th Symposium on Operating Systems Principles (SOSP)}, pages 390--405, 2017.

\bibitem{kumar2019picnic}
Praveen Kumar, Nandita Dukkipati, Nathan Lewis, Yi~Cui, Yaogong Wang, Chonggang Li, Valas Valancius, Jake Adriaens, Steve Gribble, Nate Foster, and Amin Vahdat.
\newblock Picnic: predictable virtualized nic.
\newblock In {\em Proceedings of the ACM SIGCOMM Conference}, pages 351--366, 2019.

\bibitem{kwon2020fvm}
Dongup Kwon, Junehyuk Boo, Dongryeong Kim, and Jangwoo Kim.
\newblock Fvm: Fpga-assisted virtual device emulation for fast, scalable, and flexible storage virtualization.
\newblock In {\em Proceedings of the 14th USENIX Symposium on Operating Systems Design and Implementation (OSDI)}, pages 955--971, 2020.

\bibitem{kwon2021flexcsv}
Dongup Kwon, Dongryeong Kim, Junehyuk Boo, Wonsik Lee, and Jangwoo Kim.
\newblock A fast and flexible hardware-based virtualization mechanism for computational storage devices.
\newblock In {\em Proceedings of the 2021 Annual Technical Conference (ATC)}, pages 729--743, 2021.

\bibitem{li2020leapio}
Huaicheng Li, Mingzhe Hao, Stanko Novakovic, Vaibhav Gogte, Sriram Govindan, Dan~R.K. Ports, Irene Zhang, Ricardo Bianchini, Haryadi~S. Gunawi, and Anirudh Badam.
\newblock Leapio: Efficient and portable virtual nvme storage on arm socs.
\newblock In {\em Proceedings of the 25th ACM International Conference on Architectural Support for Programming Languages and Operating Systems (ASPLOS)}, pages 591--605, 2020.

\bibitem{lim2014mica}
Hyeontaek Lim, Dongsu Han, David~G. Andersen, and Michael Kaminsky.
\newblock Mica: A holistic approach to fast in-memory key-value storage.
\newblock In {\em Proceedings fo the 11th USENIX Symposium on Networked Systems Design and Implementation (NSDI)}, pages 429--444, 2014.

\bibitem{lim2024beehive}
Katie Lim, Matthew Giordano, Theano Stavrinos, Baris Kasikci, and Thomas Anderson.
\newblock Beehive: A flexible network stack for direct-attached accelerators.
\newblock {\em arXiv preprint arXiv:2403.14770}, 2024.

\bibitem{lin2020panic}
Jiaxin Lin, Kiran Patel, Brent~E. Stephens, Anirudh Sivaraman, and Aditya Akella.
\newblock Panic: A high-performance programmable nic for multi-tenant networks.
\newblock In {\em Proceedings of the 14th USENIX Symposium on Operating Systems Design and Implementation (OSDI)}, pages 243--259, 2020.

\bibitem{liu2019ipipe}
Ming Liu, Tianyi Cui, Henry Schuh, Arvind Krishnamurthy, Simon Peter, and Karan Gupta.
\newblock Offloading distributed applications onto smartnics using ipipe.
\newblock In {\em Proceedings of the ACM SIGCOMM Conference}, pages 318--333, 2019.

\bibitem{liu2019e3}
Ming Liu, Simon Peter, Arvind Krishnamurthy, and Phitchaya~Mangpo Phothilimthana.
\newblock E3: energy-efficient microservices on smartnic-accelerated servers.
\newblock In {\em Proceedings of the USENIX Annual Technical Conference (ATC)}, pages 363--378, 2019.

\bibitem{min2021gimbal}
Jaehong Min, Ming Liu, Tapan Chugh, Chenxingyu Zhao, Andrew Wei, In~Hwan Doh, and Arvind Krishnamurthy.
\newblock Gimbal: enabling multi-tenant storage disaggregation on smartnic jbofs.
\newblock In {\em Proceedings of the ACM SIGCOMM Conference}, pages 106--122, 2021.

\bibitem{aws-gpu-video-encoding}
Macey Neff.
\newblock Gpu for video encoding optimizing video encoding with ffmpeg using nvidia gpu-based amazon ec2 instances.
\newblock \url{https://aws.amazon.com/blogs/compute/optimizing-video-encoding-with-ffmpeg-using-nvidia-gpu-based-amazon-ec2-instances/}, 2024.
\newblock Accessed: 2024-9-30.

\bibitem{neugebauer2018understandingpcie}
Rolf Neugebauer, Gianni Antichi, Jos{\'e}~Fernando Zazo, Yury Audzevich, Sergio L{\'o}pez-Buedo, and Andrew~W. Moore.
\newblock Understanding pcie performance for end host networking.
\newblock In {\em Proceedings of the ACM SIGCOMM Conference}, pages 327--341, 2018.

\bibitem{pismenny2021autonomous}
Boris Pismenny, Haggai Eran, Aviad Yehezkel, Liran Liss, Adam Morrison, and Dan Tsafrir.
\newblock Autonomous nic offloads.
\newblock In {\em Proceedings of the 26th ACM International Conference on Architectural Support for Programming Languages and Operating Systems (ASPLOS)}, pages 18--35, 2021.

\bibitem{radhakrishnan2014senic}
Sivasankar Radhakrishnan, Yilong Geng, Vimalkumar Jeyakumar, Abdul Kabbani, George Porter, and Amin Vahdat.
\newblock Senic: Scalable nic for end-host rate limiting.
\newblock In {\em Proceedings of the 11th USENIX Symposium on Networked Systems Design and Implementation (NSDI)}, pages 475--488, 2014.

\bibitem{raghavan2021breakfast}
Deepti Raghavan, Philip Levis, Matei Zaharia, and Irene Zhang.
\newblock Breakfast of champions: towards zero-copy serialization with nic scatter-gather.
\newblock In {\em Proceedings of the 18th ACM Workshop on Hot Topics in Operating Systems (HotOS)}, pages 199--205, 2021.

\bibitem{ranganathan2021google-video-accelerator}
Parthasarathy Ranganathan, Daniel Stodolsky, Jeff Calow, Jeremy Dorfman, Marisabel Guevara, Clinton~Wills Smullen~IV, Aki Kuusela, Raghu Balasubramanian, Sandeep Bhatia, Prakash Chauhan, Anna Cheung, In~Suk Chong, Niranjani Dasharathi, Jia Feng, Brian Fosco, Samuel Foss, Ben Gelb, Sarah~J. Gwin, Yoshiaki Hase, Da-ke He, C.~Richard Ho, Roy~W. Huffman~Jr., Elisha Indupalli, Indira Jayaram, Poonacha Kongetira, Cho MonKyaw, Aaron Laursen, Yuan Li, Fong Lou, Kyle~A. Lucke, JP~Maaninen, Ramon Macias, Maire Mahony, David~Alexander Munday, Srikanth Muroor, Narayana Penukonda, Eric Perkins-Argueta, Devin Persaud, Alex Ramirez, Ville-Mikko Rautio, Yolanda Ripley, Amir Salek, Sathish Sekar, Sergey~N. Sokolov, Rob Springer, Don Stark, Mercedes Tan, Mark~S. Wachsler, Andrew~C. Walton, David~A. Wickeraad, Alvin Wijaya, and Hon~Kwan Wu.
\newblock Warehouse-scale video acceleration: co-design and deployment in the wild.
\newblock In {\em Proceedings of the 26th ACM International Conference on Architectural Support for Programming Languages and Operating Systems (ASPLOS)}, pages 600--615, 2021.

\bibitem{nginx-ratelimit}
Amir Rawdat.
\newblock Rate limiting with nginx and nginx plus.
\newblock \url{https://www.nginx.com/blog/rate-limiting-nginx/}, 2017.
\newblock Accessed: 2024-8-17.

\bibitem{ruan2019insider}
Zhenyuan Ruan, Tong He, and Jason Cong.
\newblock Insider: Designing in-storage computing system for emerging high-performance drive.
\newblock In {\em Proceedings of the 2019 USENIX Annual Technical Conference (ATC)}, pages 379--394, 2019.

\bibitem{sadok2023enso}
Hugo Sadok, Nirav Atre, Zhipeng Zhao, Daniel~S. Berger, James~C. Hoe, Aurojit Panda, Justine Sherry, and Ren Wang.
\newblock Ens{\=o}: A streaming interface for nic-application communication.
\newblock In {\em Proceedings of the 17th USENIX Symposium on Operating Systems Design and Implementation (OSDI)}, pages 1005--1025, 2023.

\bibitem{shu2019dua}
Ran Shu, Peng Cheng, Guo Chen, Zhiyuan Guo, Lei Qu, Yongqiang Xiong, Derek Chiou, and Thomas Moscibroda.
\newblock Direct universal access: Making data center resources available to fpga.
\newblock In {\em Proceedings of the 16th USENIX Symposium on Networked Systems Design and Implementation (NSDI)}, pages 127--140, 2019.

\bibitem{silberstein2016gpunet}
Mark Silberstein, Sangman Kim, Seonggu Huh, Xinya Zhang, Yige Hu, Amir Wated, and Emmett Witchel.
\newblock Gpunet: Networking abstractions for gpu programs.
\newblock {\em ACM Transactions on Computer Systems (TOCS)}, 34(3):1--31, 2016.

\bibitem{son2017protego}
Jeongseok Son, Yongqiang Xiong, Kun Tan, Paul Wang, Ze~Gan, and Sue Moon.
\newblock Protego: Cloud-scale multi-tenant ipsec gateway.
\newblock In {\em Proceedings of the 2017 USENIX Annual Technical Conference (ATC)}, pages 473--485, 2017.

\bibitem{sriraman2020accelerometer}
Akshitha Sriraman and Abhishek Dhanotia.
\newblock Accelerometer: Understanding acceleration opportunities for data center overheads at hyperscale.
\newblock In {\em Proceedings of the 25th ACM International Conference on Architectural Support for Programming Languages and Operating Systems (ASPLOS)}, pages 733--750, 2020.

\bibitem{sutherland2020nebula}
Mark Sutherland, Siddharth Gupta, Babak Falsafi, Virendra Marathe, Dionisios Pnevmatikatos, and Alexandros Daglis.
\newblock The nebula rpc-optimized architecture.
\newblock In {\em Proceedings of the 47th ACM/IEEE International Symposium on Computer Architecture (ISCA)}, pages 199--212, 2020.

\bibitem{taylor2020asiccloud}
Michael~Bedford Taylor, Luis Vega, Moein Khazraee, Ikuo Magaki, Scott Davidson, and Dustin Richmond.
\newblock Asic clouds: Specializing the datacenter for planet-scale applications.
\newblock {\em Communications of the ACM}, 63(7):103--109, 2020.

\bibitem{tian2021cloudfpga-pcie-contention}
Shanquan Tian, Ilias Giechaskiel, Wenjie Xiong, and Jakub Szefer.
\newblock Cloud fpga cartography using pcie contention.
\newblock In {\em Proceedings of the 29th IEEE Annual International Symposium on Field-Programmable Custom Computing Machines (FCCM)}, pages 224--232, 2021.

\bibitem{tork2020lynx}
Maroun Tork, Lina Maudlej, and Mark Silberstein.
\newblock Lynx: A smartnic-driven accelerator-centric architecture for network servers.
\newblock In {\em Proceedings of the 25th ACM International Conference on Architectural Support for Programming Languages and Operating Systems (ASPLOS)}, pages 117--131, 2020.

\bibitem{vuppalapati2024understanding}
Midhul Vuppalapati, Saksham Agarwal, Henry Schuh, Baris Kasikci, Arvind Krishnamurthy, and Rachit Agarwal.
\newblock Understanding the host network.
\newblock In {\em Proceedings of the ACM SIGCOMM Conference}, pages 581--594, 2024.

\bibitem{wang2022fpganic}
Zeke Wang, Hongjing Huang, Jie Zhang, Fei Wu, and Gustavo Alonso.
\newblock Fpganic: An fpga-based versatile 100gb smartnic for gpus.
\newblock In {\em Proceedings of the 2022 USENIX Annual Technical Conference (ATC)}, pages 967--986, 2022.

\bibitem{wei2023characterizing}
Xingda Wei, Rongxin Cheng, Yuhan Yang, Rong Chen, and Haibo Chen.
\newblock Characterizing off-path smartnic for accelerating distributed systems.
\newblock In {\em Proceedings of the 17th USENIX Symposium on Operating Systems Design and Implementation (OSDI)}, pages 987--1004, 2023.

\bibitem{wolnikowski2021zerializer}
Adam Wolnikowski, Stephen Ibanez, Jonathan Stone, Changhoon Kim, Rajit Manohar, and Robert Soul{\'e}.
\newblock Zerializer: Towards zero-copy serialization.
\newblock In {\em Proceedings of the 18th Workshop on Hot Topics in Operating Systems (HotOS)}, pages 206--212, 2021.

\bibitem{wu2024tomur}
Shaofeng Wu, Qiang Su, Zhixiong Niu, and Hong Xu.
\newblock Tomur: Traffic-aware performance prediction of on-nic network functions with multi-resource contention.
\newblock {\em arXiv preprint arXiv:2405.05529}, 2024.

\bibitem{yu2019ava}
Hangchen Yu, Arthur~M. Peters, Amogh Akshintala, and Christopher~J. Rossbach.
\newblock Automatic virtualization of accelerators.
\newblock In {\em Proceedings of the 17th ACM Workshop on Hot Topics in Operating Systems (HotOS)}, pages 58--65, 2019.

\bibitem{zazo2015pciedma-40gbpsfpga}
Jose~Fernando Zazo, Sergio Lopez-Buedo, Yury Audzevich, and Andrew~W. Moore.
\newblock A pcie dma engine to support the virtualization of 40 gbps fpga-accelerated network appliances.
\newblock In {\em Proceedings of the 10th IEEE International Conference on ReConFigurable Computing and FPGAs (ReConFig)}, pages 1--6, 2015.

\bibitem{zhang2018gnet}
Kai Zhang, Bingsheng He, Jiayu Hu, Zeke Wang, Bei Hua, Jiayi Meng, and Lishan Yang.
\newblock G-net: Effective gpu sharing in nfv systems.
\newblock In {\em Proceedings of the 15th USENIX Symposium on Networked Systems Design and Implementation (NSDI)}, pages 187--200, 2018.

\bibitem{zhang2022justitia}
Yiwen Zhang, Yue Tan, Brent~E. Stephens, and Mosharaf Chowdhury.
\newblock Justitia: Software multi-tenancy in hardware kernel-bypass networks.
\newblock In {\em Proceedings of the 19th USENIX Symposium on Networked Systems Design and Implementation (NSDI)}, pages 1307--1326, 2022.

\bibitem{zhao2024accelerator-as-a-service}
Jiechen Zhao, Ran Shu, Katie Lim, Zewen Fan, Thomas Anderson, Mingyu Gao, and Natalie Enright~Jerger.
\newblock Accelerator-as-a-service in public clouds: An intra-host traffic management view for performance isolation in the wild.
\newblock {\em arXiv preprint arXiv:2407.10098}, 2024.

\bibitem{zhao2024smartniclivemigration}
Jiechen Zhao, Ran Shu, Lei Qu, Ziyue Yang, Natalie Enright~Jerger, Derek Chiou, Peng Cheng, and Yongqiang Xiong.
\newblock Smartnic-enabled live migration for storage-optimized vms.
\newblock In {\em Proceedings of the 15th ACM SIGOPS Asia-Pacific Workshop on Systems (APSys)}, pages 45--52, 2024.

\bibitem{zhao2022altocumulus}
Jiechen Zhao, Iris Uwizeyimana, Karthik Ganesan, Mark~C. Jeffrey, and Natalie Enright~Jerger.
\newblock Altocumulus: Scalable scheduling for nanosecond-scale remote procedure calls.
\newblock In {\em Proceedings of the 55th IEEE/ACM International Symposium on Microarchitecture (MICRO)}, pages 423--440, 2022.

\end{thebibliography}
